\documentclass[lettersize,journal, multi, tikz]{IEEEtran} %lettersize,journal
\usepackage{amsmath,amsfonts}
\usepackage{algorithmic}
\usepackage{algorithm}
\usepackage{array}
\usepackage[caption=false,font=normalsize,labelfont=sf,textfont=sf]{subfig}
\usepackage{textcomp}
\usepackage{stfloats}
\usepackage{url}
\usepackage{verbatim}
\usepackage{graphicx}
\usepackage{cite}
\usepackage{enumitem}
\usepackage{multirow}

\usepackage[affil-it]{authblk} 
\begin{document}

\title{Bridging the Spoof Gap: A Unified Parallel Aggregation Network for Voice Presentation Attacks}

\author[1]{Awais Khan}
\author[1]{Khalid Mahmood Malik~\IEEEmembership{(Senior Member,~IEEE)}}
\affil[1]{Department of Computer Science and Engineering, Oakland University, Rochester, MI 48309, USA}
        % <-this % stops a space
%\thanks{This paper was produced by the IEEE Publication Technology Group. They are in Piscataway, NJ.}% <-this % stops a space
%\thanks{Manuscript received April 19, 2021; revised August 16, 2021.}

\markboth{Journal of \LaTeX\ Class Files,~Vol.~14, No.~8, August~2021}%
{Shell \MakeLowercase{\textit{et al.}}: A Sample Article Using IEEEtran.cls for IEEE Journals}


\maketitle

\begin{abstract}
Automatic Speaker Verification (ASV) systems are increasingly used in voice biometrics for user authentication but are susceptible to logical and physical spoofing attacks, posing security risks. Existing research mainly tackles logical or physical attacks separately, leading to a gap in unified spoofing detection. Moreover, when existing systems attempt to handle both types of attacks, they often exhibit significant disparities in the Equal Error Rate (EER). To bridge this gap, we present a Parallel Stacked Aggregation Network that processes raw audio. Our approach employs a split-transform-aggregation technique, dividing utterances into convolved representations, applying transformations, and aggregating the results to identify logical (LA) and physical (PA) spoofing attacks. Evaluation of the ASVspoof-2019 and VSDC datasets shows the effectiveness of the proposed system. It outperforms state-of-the-art solutions, displaying reduced EER disparities and superior performance in detecting spoofing attacks. This highlights the proposed method's generalizability and superiority. In a world increasingly reliant on voice-based security, our unified spoofing detection system provides a robust defense against a spectrum of voice spoofing attacks, safeguarding ASVs and user data effectively.
\end{abstract}


\section{Introduction}
\IEEEPARstart{B}{iometrics}, a vital progression from traditional password-based authentication, are gaining widespread adoption for user identification across various applications. In recent years, Automatic Speaker Verification (ASV), a type of voice biometrics, has gained prominence for its ability to authenticate users based on unique speech characteristics. The ASVs are also experiencing growing utilization in smart speakers (such as Google Home, Siri, Amazon Alexa) and various Internet of Things (IoT) devices, enabling voice-activated access to services and resources \cite{obaidat2019biometric}. 
However, despite offering cost-effective authentication, ASV systems exhibit vulnerabilities to both physical and logical voice presentation attacks (VPAs), commonly known as voice spoofing attacks. These vulnerabilities present challenges to the widespread adoption of ASV technology. 

A convenient way to deter a VPA is through the use of anti-spoofing or presentation attack detection (PAD) systems, which perform acoustic characterization of genuine
and spoofed speech signal \cite{sahidullah2019introduction}. Integrating an independent spoofing countermeasure, or PAD, with an ASV system has been demonstrated to enhance resilience against spoofing attacks \cite{khan2023battling}. Consequently, substantial research effort has been directed toward developing spoofing countermeasures, particularly targeting the four main VPA classes: impersonation, speech synthesis (SS), voice conversion (VC), and replay. While impersonation attacks, lacking standardized databases, have received comparatively less research attention, this study focuses on the remaining three presentation attack types.

\begin{figure}[!t]
        \centering
        \includegraphics[width=8cm]{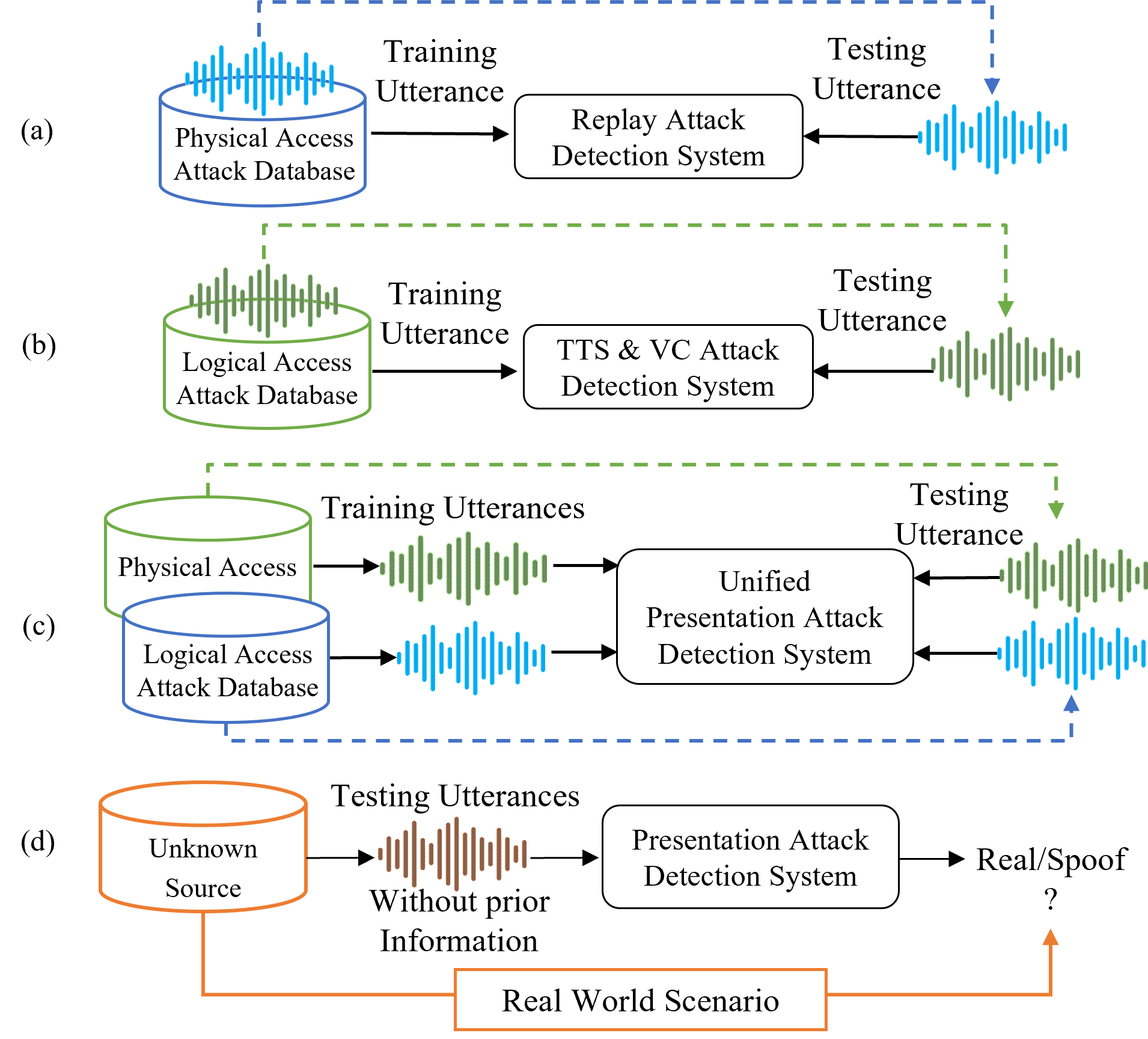}
        \caption{The architectural framework of existing anti-spoofing systems: (a) a standard replay attack detection system that counters only physical attacks (b) A standard anti-spoofing system for synthesis speech samples that was trained and tested using Logical Access speech samples (c) Unified PAD systems, trained and tested for LA and PA attacks, also trained and tested in parallel with PA and LA data samples. (d) A real-world challenge for the PADs. }
        \label{fig:UnifiedVsSingular}
\end{figure} 

In the realm of voice spoofing, various strategies have been developed to combat voice spoofing attacks, as illustrated in Figure~\ref{fig:UnifiedVsSingular}. Anti-spoofing research highlights significant differences between Speech Synthesis (SS), Voice Conversion (VC), and replay spoofing attacks~\cite{das2020assessing, khan2023battling}. SS and VC attacks use modern AI-algorithms, resulting in machine-generated distortions like robotic voices and the absence of natural pauses etc.
In contrast, replay attacks involve recorded variations of genuine speech, leading to microphonic disparities.
Unlike VC and SS, replay attack indicators often lie in the high-frequency region of the recordings~\cite{witkowski2017audio}. Due to these distinct artifact characteristics, existing countermeasures mostly target specific attack types, as shown in Figure~\ref{fig:UnifiedVsSingular} (a) and (b), with limited attention to unified spoofing detectors (Figure~\ref{fig:UnifiedVsSingular} (c)).
In practical scenarios, the nature of an attack on an ASV system is often unknown. Consequently, there is a need for unified spoof detection models capable of addressing all types of spoofing attacks and identifying unique artifacts for each specific attack. Unfortunately, there remains a significant gap in comprehensive research addressing state-of-the-art (SOTA) spoofing attacks.

In anti-spoofing literature, existing solutions can be classified into two categories: front-end features coupled with a back-end classifier and DNN-based end-to-end solutions. Previous studies have demonstrated the effectiveness of manually crafted features combined with a back-end classifier in detecting Speech Synthesis (SS) and Voice Conversion (VC) attacks, as these distortions manifest across various speech frames~\cite{sahidullah2019introduction,khan2023battling}. This success is attributed to the inclusion of specific sub-band extractions of acoustic cues within the front-end features. In contrast, distinguishing Replay (PA) attacks poses challenges for front-end features due to their utilization of a broader spectrum band that encompasses the entire utterance. Moreover, the presence of identical microphonic variations complicates the differentiation between replay attacks and legitimate speech~\cite{witkowski2017audio}. Consequently, many feature-based systems have been specifically designed to target either SS/VC (LA) attacks or replay (PA) attacks exclusively.

Following recent advancements, the emphasis of anti-spoofing research has shifted from "crafted features" to "end-to-end networks"~\cite{zhang122021effect,jung2022aasist,tak2021end}. However, there remains a significant variation in EERs across existing solutions, particularly when addressing the detection of logical access and physical access attacks with a single system~\cite{8272715}. For instance, in ASSERTS~\cite{lai2019assert}, the EER is reported as $0.59\%$ for PA but increases to $6.70\%$ for LA. Similarly, in STC~\cite{lavrentyeva2019stc}, the EER stands at 4.6\% for PA and 7.86\% for LA, while comparable performance variations are observed in BUT-Omilia~\cite{zeinali2019detecting}, MFMT~\cite{li2019anti}, and SASV~\cite{aljasem2021secure} solutions. These results shows the bias of existing systems towards either LA or PA attacks and raise concerns about the practicality of deploying existing PADs in real-world scenarios.

Furthermore, with the exception of the system in \cite{tak2021end}, most of the unified systems rely on frontend features or spectrogrammatic representations of speech samples for input. This dependence on resource-intensive computation raises concerns about the applicability of the system to resource-constrained edge devices. In parallel, this demonstrates the research gap when it comes to the direct use of speech signals to distinguish spoof from real utterances. In particular, this paper attempt to answer the following research questions:
\begin{itemize}
    \item Are the unified detectors equally good at detecting both logical and physical attacks? Does the proposed aggregated network show better cumulative EERs for LA and PA compared to state-of-the-art unified solutions using raw audio signal?
    \item How do existing residual networks perform in terms of EER, and is there any need to use aggregated networks? 
    \item What is the trade-off between model widths and density in aggregated networks? Further, what density and width are optimal for a unified anti-spoofing system? 
    \item What input and DNN model is optimal for resource-constrained devices when integrating PADs into ASV systems? And what is the performance of the aggregated models compared to existing handcrafted features?
\end{itemize}
To answer the above questions, we present a parallel stack aggregated (PSA) network, as illustrated in \cite{xie2017aggregated}. The PSA network leverages the split-transform-merge strategy from Inception networks for effective extraction of frame-level acoustic cues. Simultaneously, it employs the repeating residual layer architecture from VGG-Net and ResNets in a scalable manner to capture utterance-level speech representations. Our network collectively applies a series of transformations to a low-dimensional embedding, facilitating the extraction of both frame-level and utterance-level representations. These outputs are then aggregated to obtain finely detailed acoustic representations, subsequently processed in a dense layer architecture to discriminate between spoof and authentic speech. Moreover, rather than transforming the waveform from the time domain to the frequency domain and then developing classifiers, the proposed system learns all at once from the raw speech signal. To sum up, the main contributions of this work are as follows:
\begin{enumerate}
    \item We introduce a unified SE-Parallel Stacked Aggregated Network designed to detect a range of speech presentation attacks using raw audio data, without being constrained by the computational complexities of spectrograms or manually crafted features.
    \item We examine the efficacy of several residual and aggregated residual networks using squeeze and excitation (SE) networks. To the best of our knowledge, we are the first to use SE in concert with a parallel stacked aggregated network to use raw audio in order to combat LA and PA voice presentation attacks. 
    \item The presented system surpasses twelve individual and seven unified solutions, including the baseline models used in the ASVspoof2019 challenge. It notably mitigates the bias observed in state-of-the-art unified solutions towards a particular attack class.
    \item  We conduct a thorough ablation study using eight networks to combat advanced voice presentation attacks. Our system outperforms both comparative models and ASVspoof2019 baseline systems across standalone and unified testing across two datasets.
\end{enumerate}{}{}

The rest of the paper is organized as follows: Section 2 reviews prior work in the field, while Section 3 elaborates on the SE-PSA network development methodology. Section 4 outlines the experimental setup, and Section 5 examines the results, including comparisons with other systems, as well as the ablation study. Finally, Section 6 offers the conclusion.

\section{Existing Work}
Existing research on voice anti-spoofing solutions falls into two main categories: hand-crafted countermeasures with frame-level classifiers such as GMM and i-vectors~\cite{khan2022voice}, and DNN-based end-to-end solutions~\cite{li2021replay}.
\subsection{standalone systems for anti-spoofing}
Most spoofing research primarily focuses on developing standalone anti-spoofing systems, which commonly utilize handcrafted features and backend classifiers. These techniques vary mainly in the types of features and classifiers employed. Various features, such as magnitude-spectrum-based~\cite{wu2012detecting, sriskandaraja2016front}, phase-spectrum-based~\cite{yang2021modified}, and modulation-spectrum-based features~\cite{kamble2020amplitude}, have been explored in combination with classifiers like Gaussian mixture models~\cite{sahidullah2015comparison, khan2022voice, khan2022toward} and support vector machines~\cite{rahmeni2020speech}. While these approaches have advanced attack identification, they often involve prior task-specific speech manipulation and short-term spectral auditory processing.

With the rise of neural networks, the research community has shifted its focus toward integrating front-end features with neural network architectures, including CNN~\cite{albawi2017understanding}, ResNet~\cite{li2021replay}, and attention-based methods~\cite{yu2018deep}. This has led to the development of numerous anti-spoofing systems that combine advanced DNNs with handcrafted features to capture more discriminative local descriptors~\cite{jung2022aasist, zhang122021effect, tak2021end}. Although these systems have significantly outperformed traditional machine-learning-based countermeasures, such as RW-Resnet~\cite{ma2021rw}, SENET~\cite{zhang122021effect} and~Assist \cite{jung2022aasist}, none have been designed to address both LA and PA attacks within a single system.
\subsection{Unified solutions and their limitations}
In \cite{lavrentyeva2019stc}, the author presents a unified anti-spoofing system based on numerous front-end features. The authors found that the system that used LFCC performed better overall than other features. However, failed to detect synthesized speech effectively, and its EER remained much higher when presented with replay attacks. Similarly, Li et al.~\cite{li2019anti} employed multi-task learning with multi-feature integration (MFCC, CQCC, and Filter Bank) to detect both LA and PA spoofing attacks. While this combination of cepstral features performed well for replay attack detection (EER of $0.96\%$), it struggled with LA attacks. In contrast to the cepstral features, a combination of ternary features, named sm-ALTP \cite{aljasem2021secure}, was presented to detect voice spoofing. Even after suppressing the performance of cepstral features with aggregation-based ensemble, the ternary features struggled against LA attacks \cite{aljasem2021secure}. In another study, Zeinali et al. \cite{zeinali2019detecting} presented an anti-spoofing system that merged two VGG networks trained on single- and two-feature sets. While it performed reasonably well with power spectrogram and CQT features, it encountered higher EERs in LA attack testing. In the ASSERTS system \cite{lai2019assert}, CQCC features and a squeeze-and-excitation-based residual network were used to identify VS, SS, and replay attacks. While the model excelled in replay detection with a low $0.59\%$ EER, it struggled to identify SS and VC attacks, resulting in an increased EER of $6.70\%$ during evaluation. Thus, both standalone and unified spoofing countermeasures exhibited significant EER disparities in the detection of SS/VC versus replay attacks. Additionally, each solution heavily relied on specifically extracted features or spectrograms.

\begin{figure}[!b]
        \centering
        \includegraphics[width=8cm]{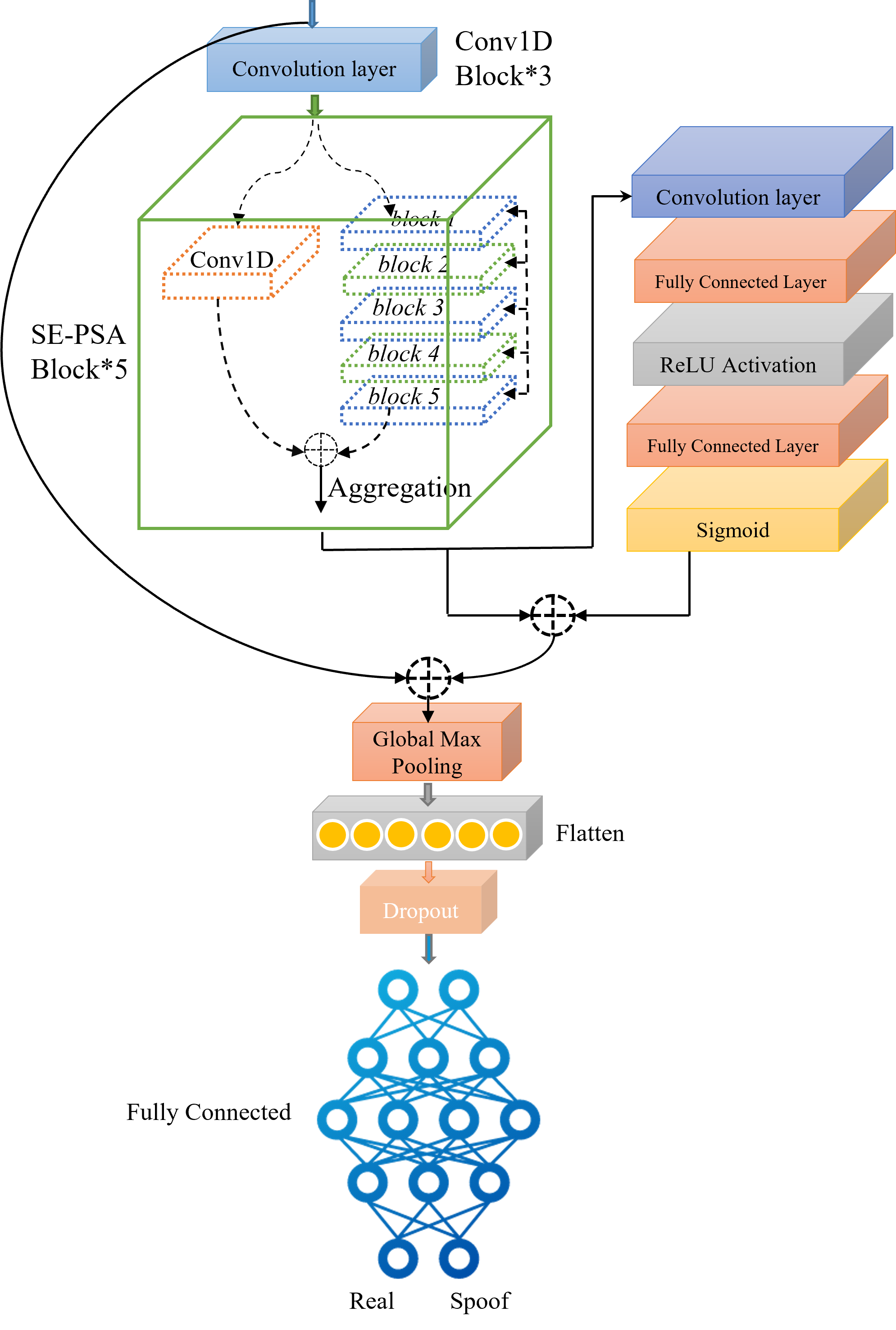}
        \caption{The internal architectural framework for addressing gradient vanishing via spatial dropout. (a) A standard DNN with processing and activation of all neurons without any selection or drop (b) A standard neural network with spatial dropout, which causes the selection of required neurons with more crucial embeddings\cite{lee2020revisiting}.}
        \label{fig:overallFramework}
\end{figure}
\section{Methodology}
This section provides an overview of the proposed SE-PSA anti-spoofing system as well as details on each phase. The proposed anti-spoofing system is divided into two stages: data preparation (composed of data pre-processing and augmentation presented in Section IV) and the parallel stacking aggregate network. 

\subsection{Parallel Stack Aggregation Network}
The Parallel Stack Aggregated (PSA) network follows a topology identical to ResNeXt's intra-architecture \cite{xie2017aggregated}. The design in \cite{xie2017aggregated} has shown that it appears to reduce the risk of hyper-parameter over-adaptation to a specific dataset, which is currently missing in existing ResNet models. While aggregated network architectures have proven effective in image classification, their application in audio classification or spoofing detection remains unexplored. This paper addresses the challenge of creating a unified solution for detecting various audio forgeries by integrating the Split-Transform-Merge (STM) strategy and squeeze and excitation approaches. Additionally, it incorporates a stacking-based classifier capable of effectively discerning artifacts in both genuine and spoofed speech samples. Notably, the STM approach achieves these objectives with reduced computational complexity, aligning with our goal of providing a lightweight solution. This approach works directly on raw waveforms, eliminating the need for additional spectrogram generation or handcrafted feature extraction. 

The PSA Network comprises two main components: the first component involves passing the pre-processed speech signal through convolution blocks to extract convolved embeddings. These embeddings are then forwarded to the second component, the SE-PSA blocks, which utilize them to extract fine-grained features necessary for classifying spoofed and authentic speech samples. Specifically, the SE-PSA blocks employ a structured VGG/ResNets architecture, combining the repeating strategy of ResNet with the split-transform-merge strategy of the Inception Network. Each network block divides the input, transforms it as required, and aggregates it to generate the output. All blocks within the network follow this parallel topology. Figure \ref{fig:overallFramework} provides a visual representation of both the overall and internal architecture of the PSA network. The convolution block consists of three convolution layers for extracting convolved embeddings, while the five SE-PSA blocks comprise four cardinal paths with group convolutions. The final block serves the purpose of classifying speech samples as either genuine or spoofed.

In the initial stage, the input audio signal denoted as $F[n]$, consisting of $N$ samples corresponding to frames $n = {1, 2, 3, \ldots, k}$ containing spectral and temporal details, is fed into the first convolutional block of the network. This convolution block comprises three layers with $c1, c2,$ and $c3$ filters, $k1, k2,$ and $k3$ kernel sizes, strides, and employs the same padding, along with a softmax activation function. It has been observed that pre-activation convolution yields better results in voice spoofing detection compared to post-activation convolution. Therefore, to generate a deep feature map of the speech signal, we utilize a pre-activation convolution block that includes batch normalization, activation, followed by a convolution layer. Once the deep feature map is obtained, max pooling is applied to extract the enhanced embedding $ E^{st}_c = {e_1, e_2, e_3, \ldots, e_n}$ representing the speech signal. Detailed specifications regarding convolution size, strides, and filter usage for extracting discriminative embedding representations are provided in Table \ref{tab:PSA}.
\begin{table}[b]
\caption{Detailed architecture of the PSA Network. Adaptive average pooling and global average pooling are selected as pooling layers, and the output size is set to (1,1), which results in each channel having exactly one output to feed to the fully connected layers.}
\label{tab:PSA}
\centering/
\begin{tabular}{c | p{1.5cm}  | p{2.4cm} | c }
  \hline
  Layer Name & Output Size & SE-PSA  & Channel  \\
  \hline
 \multirow{2}{1cm}{Conv1D} & \multirow{2}{2cm}{$T \times F $ }& \multirow{1}{2.2cm}{$7 \times 7$ , 64, stride 2}   & \multirow{2}{0.4cm}{16} \\ 
           &           &        $max pool, stride 2 $    &   \\ \hline
 \multirow{2}{1cm}{SE-PSA} &\multirow{2}{2cm}{$T \times F$ } & \multirow{1}{2.2cm}{$ \left[ 3\times 3,32\right] \times 2 $}  & \multirow{2}{0.4cm}{32}  \\
           &   & $\left[ 3\times 3,32 \right],  C=4 $  &    \\ \hline
\multirow{2}{1cm}{SE-PSA} &\multirow{2}{2cm}{$T/2 \times F/2$} & \multirow{1}{2.2cm}{$ \left[ 3\times 3,64\right] \times 2 $}  & \multirow{2}{0.4cm}{64} \\
           &   & $\left[ 3 \times 3,64 \right], C=4 $  &     \\ \hline
 \multirow{2}{1cm}{SE-PSA} & \multirow{2}{2cm}{$T/4 \times F/4$} & \multirow{1}{2.2cm}{$ \left[ 3\times 3,128\right] \times 2 $} & \multirow{2}{0.6cm}{128} \\
     &  & $\left[ 3 \times 3,128 \right], C=4$ &    \\ \hline
\multirow{2}{1cm}{SE-PSA} & \multirow{2}{2cm}{$T/4 \times F/4$}&\multirow{1}{2.2cm}{$ \left[ 3\times 3,256\right] \times 2 $}& \multirow{2}{0.6cm}{256} \\
      &   & $\left[ 3 \times 3,256 \right], C=4 $  &    \\ \hline
\multirow{2}{1cm}{SE-PSA} & \multirow{2}{2cm}{$T/8 \times F/8$ }& \multirow{1}{2.2cm}{$\left[ 3\times 3,512\right] \times 2 $} & \multirow{2}{0.6cm}{512} \\
        &           & $\left[ 3\times 3,512 \right], C=4  $  &   \\ \hline
 \multirow{2}{1cm}{$Output$} &  \multirow{2}{2.5cm}{$1 \times 1$}  &  $Global Max Pool$, 1000-d fc, sigmoid  & \multirow{2}{0.6cm}{$--$ }\\ \hline
$FLOPs$  & $1.8 \times 10^{9}$ & \multirow{1}{2cm}{$---$} & $---$ \\
 \hline      
\end{tabular}
 \end{table}

\begin{figure*}[!t]
        \centering
        \includegraphics[width=12cm, height= 10cm]{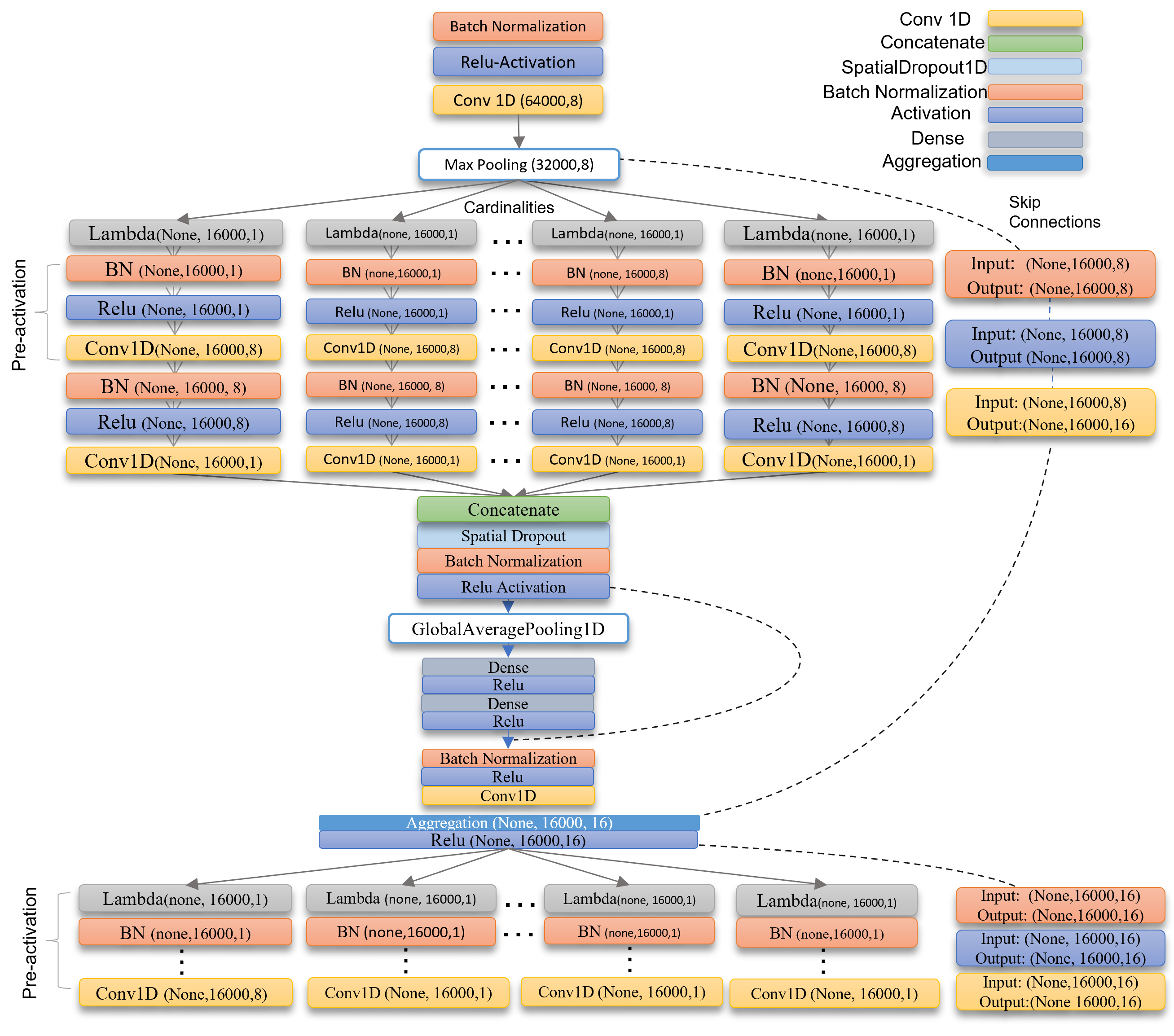}%{Intra-architecture.png}
        \caption{Intra-architecture of SE-PSA Blocks with 4 cardinalities and pre-activation convolutions. The similar Intra- architecture repeated 5 times for each block of the proposed PSA network.}
        \label{fig:Intra-architecture}
\end{figure*}
In the second fold, the acquired embedding $ E^{st}_c$ is passed to the SE-PSA block, responsible for extracting fine-grained representations denoted as $ F^{fg}_r = {f^{g}_1, f^{g}_2, f^{g}_3, \ldots, f^{g}_n}$ used for the classification of genuine and spoofed speech samples.

The core architecture of the PSA block adheres to the same two principles governing the ResNeXt architecture:
\begin{itemize}
    \item When creating spatial maps of identical size, blocks share identical hyper-parameters (width and filter sizes).
\item The width of the blocks is increased by a factor of $2$, leading to an increase in width each time the spatial map is down-sampled by a factor of $2$.
\end{itemize}
These principles significantly streamline the design and enable us to focus on a few critical factors. In the intra-architecture illustrated in Fig. \ref{fig:Intra-architecture}, the PSA network combines the high-level $H{f^st}$ and low-level feature $L{f^st}$ representations extracted from homogeneous neural paths. Subsequently, the high and low-level feature representations obtained from this step are passed on to the next SE-PSA block. This process is iterated $M$ times to derive an adaptive feature representation for both spoofed and legitimate speech samples.

\begin{figure}[!t]
        \centering
        \includegraphics[width=8cm]{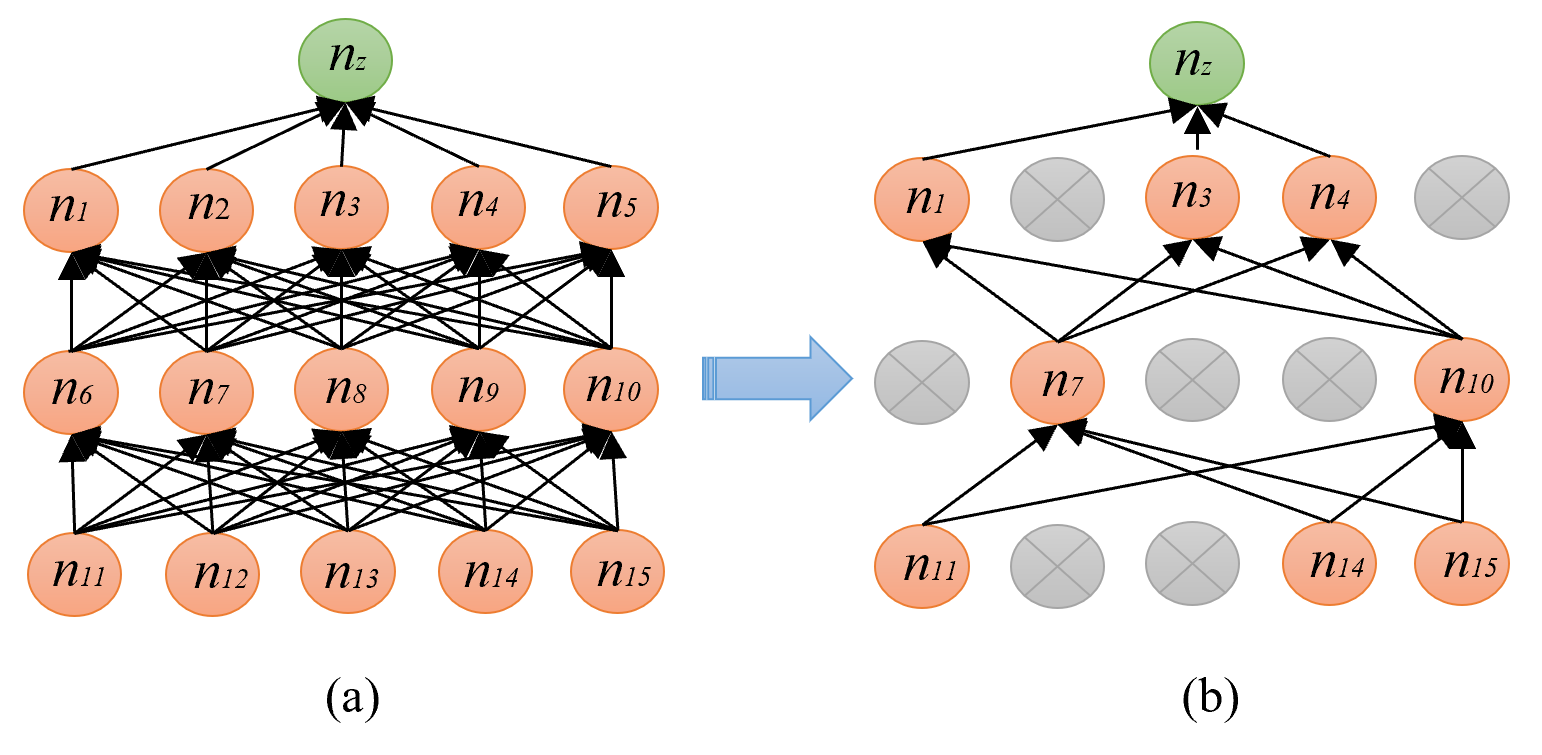}
        \caption{The internal architectural framework for addressing the vanishing gradient via Spatial dropout. (a) A standard DNN, with processing and activation of all neurons, without any selection or drop. (b) A standard neural network with Spatial dropout, which results in the selection of required neurons with more relevant embeddings\cite{lee2020revisiting}.}
        \label{fig:spatialdropout}
\end{figure}

This adaptive feature representation is achieved through the introduction of "cardinality," denoted as $C$, which adds an additional dimension to residual networks, making them wider rather than deeper. The value of cardinality ${C}$ determines the size of the transformation set $T = {t_1, t_2, \ldots, t_n}$. The same transformations $T$ are applied $M$ times, and the cumulative gain is aggregated as shown in the equations below.
\begin{equation}
    S_R = \sum^{D}_{i=1}\omega_in_i
\end{equation}
where $n_i = \{n_1, n_2..n_i$ is the $D$ channel input vector to the neuron, and $\omega_in_i$ is the filter weight for the $i_{th}$ channel. $S_R$ shows the inner product of the neural network.
\begin{equation}
    E^{st}_c = \sum^{C}_{i=1} \tau_i(n_i)
\end{equation}
\begin{equation}
    F^{fg}_r =  E^{st}_c + \sum^{C}_{i=1} \tau_i(n_i)
\end{equation}
 Where $E^{st}_c$ represents the aggregates transformation and $\tau_i(n_i)$ can be an arbitrary function. Analogous to a $S_R$, $\tau_i(n_i)$ projects 
$E^{st}_c$ embedding and then transforms and aggregated to $F^{fg}_r$. where $F^{fg}_r$ refers to the fine-grained representation extracted to classify the speech representations. Lastly, we employ global max pooling, flatten the extracted representations $F^{fg}_r$, and add the fully connected dense layers, followed by dropout to classify the real and spoofed speech samples. For the classification, we used the sigmoid activation function to extract the score of an utterance being forged or bona fide, as shown below:
\begin{equation}
   \mathbb{S}_{cr}= \frac{1}{1+e^{-x}}
\end{equation}
\begin{equation}
    \mathbb{P}_{pred}=\mathbb{S}_{cr} > 0.5
\end{equation}
Where $S_{cr}$ denotes the score of the utterance and $\mathbb{P}_{pred}$ refers to the prediction of the model as spoof and bonafide speech.
Hyper-parameters, including width and filter sizes, are shared within the SE-PSA block. Group convolutions are utilized as an aggregation strategy. To address network overfitting, spatial dropout is applied before the aggregation node. Spatial dropout involves selecting neurons with more significant embeddings and dropping those with less representative features, as illustrated in Fig. \ref{fig:spatialdropout}.

\begin{figure}[!b]
        \centering
        \includegraphics[width=8cm, height=5.3cm]{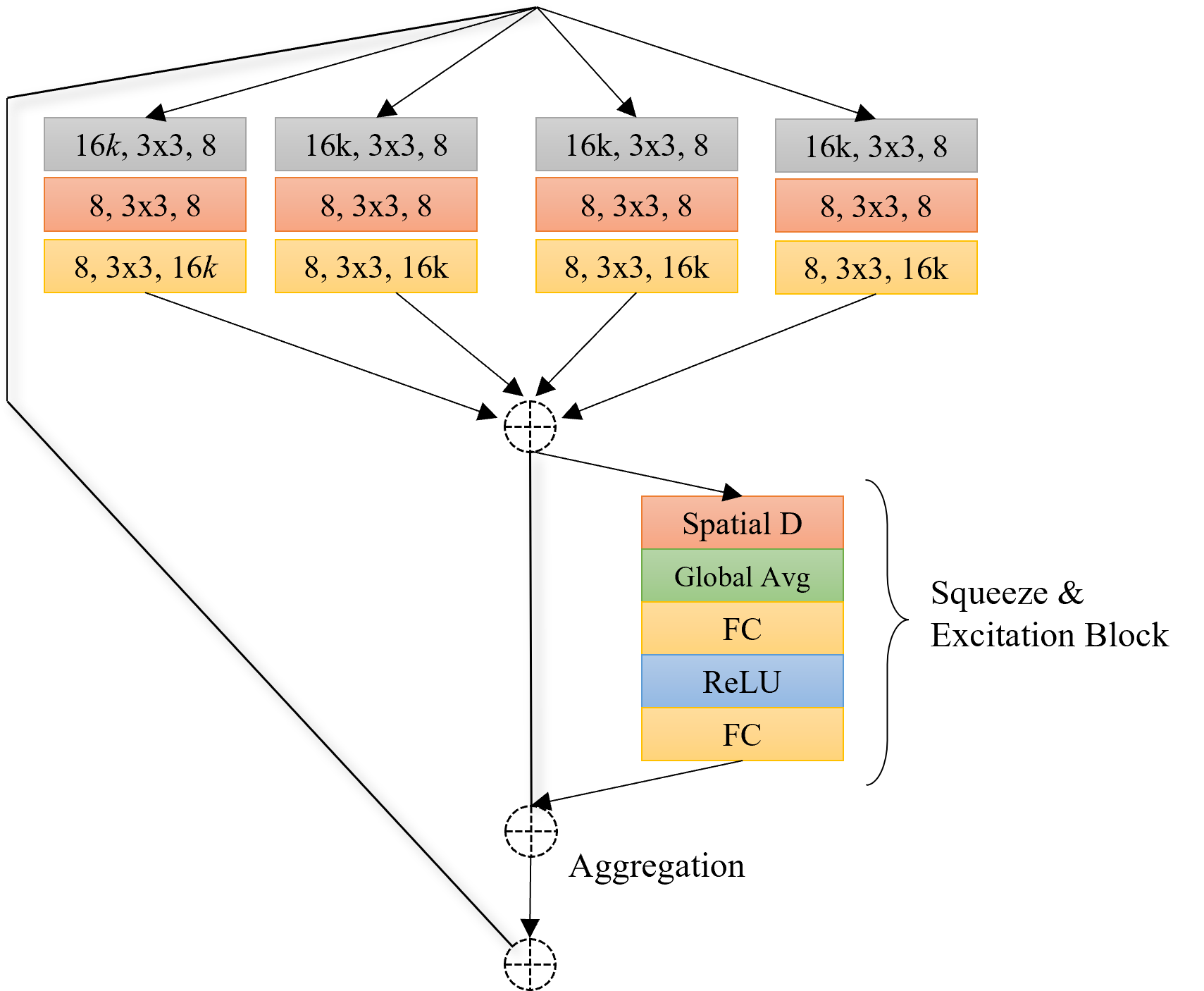}
        \caption{Aggregated Feature Map extraction with the Squeeze and Excitation Block. The SE block include the spatial dropout applied before every global average layer of the each SE-PSA block.}
        \label{fig:Squeeze&Excitation}
\end{figure}

Furthermore, recognizing the significant differences between LA and PA speech samples, we introduce Squeeze and Excitation (SE) to the PSA block. This addition effectively extracts transformed feature maps and aids in back-propagation within the network. Fig. \ref{fig:Squeeze&Excitation} illustrates the layered architecture of the SE connections.
Multi-cardinality transforms accentuate critical parameters for distinguishing between genuine and spoofed tasks, while reducing the significance of highly correlated parameters. This, combined with the residual block featuring split-transform and merge techniques, enables effective differentiation between authentic and spoofed speech.

\subsubsection{Addressing the gradient vanishing}
Skip connections are utilized to mitigate the vanishing gradient issue by maintaining the error gradient during back-propagation. They multiply the error gradient by one during back-propagation through the skip connection. This allows for the training of deeper networks, enhancing the ability of our PAD system to discern authentic from spoofed speech. In a network without skip connections, the gradient is computed as follows:
\begin{equation}
    y = \frac{\partial J}{\partial x}
\end{equation}
where $y$ obtained using the chain rule for the full operation with all the steps. Whereas in the case of without the skip connection, the full operations can be performed as shown in Eq. below:
    \begin{equation}
 \frac{\partial J}{\partial x_0} =  \frac{\partial J}{\partial x_2} \frac{\partial x_2}{\partial z_2} \frac{\partial z_2}{\partial x_1} \frac{\partial x_1}{\partial z_1}\frac{\partial z_1}{\partial x_0}
\end{equation}
This denotes the chain of multiplications which renders neural networks prone to disappearing and exploding gradients. If we substitute F(x) for the intermediate computations, the gradient calculation becomes: 
 \begin{equation}
 \frac{\partial J}{\partial x} =  \frac{\partial J}{\partial F(x)} \frac{\partial F(x)}{\partial x} 
\end{equation}
Next, we add the new function $H(x) = F(x) + x$ for the added the skip connection. In particular, we must now differentiate $F(x)$ through $H(x)$ to get the gradient of the cost function in a network with skip connections as shown below: 
 \begin{equation}
 \frac{\partial J}{\partial x} =  \frac{\partial J}{\partial H(x)} \frac{\partial F(x)}{\partial x} 
\end{equation}
where the derivative of $x$ with respect to $ H(x)$ is equal to 1. Thus, substituting $F(x) + x$ for $H(x)$ yields the expression: 
 \begin{equation}
\frac{\partial J}{\partial x} = \frac{\partial J}{\partial H(x)} ({\frac{\partial F(x)}{\partial x}} + 1) =  \frac{\partial J}{\partial H(x)} \frac{\partial F(x)}{\partial x} + \frac{\partial J}{\partial H(x)} \end{equation}
In this scenario, the gradient of $F(x)$ becomes extremely small as a result of multiple matrix multiplications during back-propagation through all the layers of $x$. However, we still retain the direct gradient of the cost function concerning $H(x)$.
This approach allows the network to bypass certain gradient computations during back-propagation, preventing gradient vanishing or exploding.
In the next section, we outline the experimental setup for conducting experiments and comparative analyses of the proposed system.

\section{Experimental setup}
In this section, we illustrate the experimental configuration used to produce the reported results. The proposed anti-spoofing system was validated against TTS, VC, replay and chained replay spoofing samples. Following are the metrics, datasets, and hyper-parameters used for all of the testing and results presented below.
\subsection{Dataset}
Since 2015 the ASVspoof challenge has been providing datasets and standards in order to promote the development of spoofing countermeasures. Among these datasets, the ASVspoof 2019 database has become the de facto standard for the investigation and evaluation of voice spoofing countermeasures. Consequently, we evaluated the performance of the proposed anti-spoofing system using the ASVspoof2019 \cite{todisco2019asvspoof}. To contrast the performance of the proposed system against single- and multi-order replay attacks we also employed the voice spoofing detection corpus (VSDC) developed in  \cite{baumann2021voice}.

The ASVspoof2019 dataset \cite{todisco2019asvspoof} comprises audio samples recorded at a 16 kHz sample rate with 16-bit compression. This dataset is divided into two categories: Logical Access (LA) and Physical Access (PA), each further divided into training, development, and evaluation subsets. Both the training and development sets contain speech samples from 20 distinct speakers, with spoofed speech samples generated using algorithms A01 to A06. In contrast, the evaluation set includes bonafide speech samples from 67 speakers and spoofed samples generated using 19 algorithms, including GANs and DNNs. For dataset details, please refer to Table \ref{tab:Asvspoof2019}, and further configuration specifics can be found in \cite{todisco2019asvspoof}.
\begin{table}[b]
\caption{A summary of ASVspoof Challenge 2019 database.}
\label{tab:Asvspoof2019}
%\centering
\begin{tabular}{p{1.5cm}|p{0.5cm}|p{0.7cm}|p{0.8cm}|p{0.8cm}|p{0.8cm}|p{0.8cm}}
  \hline
  & \multicolumn{2}{l|}{Speaker} & \multicolumn{2}{l|}{LA Attacks}  & \multicolumn{2}{l}{PA Attacks}  \\  
  \hline
 Subset & Male  & Female & Genuine & Spoofed & Genuine & Spoofed \\
 \hline
 Training & 8 & 12 & 2580 & 22800 & 5400 & 48600\\
 Development & 8 & 12 & 2548 & 22296 & 5400 & 24300\\ \hline
 Evaluation & - & - & \multicolumn{2}{l|}{71747} & \multicolumn{2}{l}{137457}  \\
 \hline
     \end{tabular}
 \end{table}
 
The system's performance against replay attacks was evaluated using VSDC \cite{baumann2021voice}, which includes first-order and second-order replay spoof samples alongside genuine speech. The dataset introduces variations in environments, configurations, genres, recording and replay devices, and output devices (speakers). In contrast to ASVspoof2019, VSDC incorporates noise and microphonic differences in speech samples and employs multiple playback devices to minimize bias. VSDC comprises 19 voices (10 male and 9 female), with each audio sample lasting 6 seconds. Table \ref{tab:VSDC} provides the details of the VSDC dataset, and \cite{baumann2021voice} contains information about the playback devices and development architecture used in its construction. 
\begin{table}[t]
\caption{A summary of VSDC Database.}
\label{tab:VSDC}
\centering
\begin{tabular}{p{2.0cm} | p{1.5cm} | p{2.5cm} | p{1.0cm}}
  \hline
  Audio samples & Speech Samples  & Environment & Sample Rate\\
 \hline
 Bonafide & 4000 & \multirow{4}{2.5cm}{Recording Chamber, Kitchen Table, Living Room, Office Desk} & \multirow{4}{4cm}{96K}\\
 Replay & 4000 &  & \\
 Clonded Replay & 4000 &  &  \\
 Total & 12000 &  & \\
 
 \hline      
\end{tabular}
 \end{table}
\subsection{Data Preprocessing}
Raw audio, characterized by discrete high- and low-frequency values and amplitude variation, requires preprocessing before input into the PSA network due to the significant data point differences. This preprocessing includes normalization and frame length adjustment. In cases like the ASVspoof2019 dataset, where audio files have variable-length segments and varying frame counts, input voice sample length ($L$) is standardized to four seconds either by concatenation or trimming. The first four seconds (equivalent to $1 \times 64000$ samples) of the speech sample are retained, and sequence padding is applied to shorter speech samples. Subsequently, Z-score normalization is conducted on the raw waveform, constraining sample values to the range $[-1, 1]$ using mean and standard deviation values, as shown below:
\begin{equation}
    x = \frac{\sum_{i=1}^{j} (x - \mu)}{\sigma} 
\end{equation}
\begin{equation}
    \sigma = \sqrt{E[X^2]-(E[X])^2} \\
\end{equation}
where $x$ is the number of speech samples, $mu$ and $sigma$ are the mean and standard deviation of the signal, and $E[X2]$, $(E[X])2$ are the mean of the squared data and the square of the mean of the data, respectively. After standardizing the data, we apply five types of augmentation to address data imbalances. The details of data augmentation are explained in the section below.
\begin{figure}[!b]
        \centering
        \includegraphics[width=8cm]{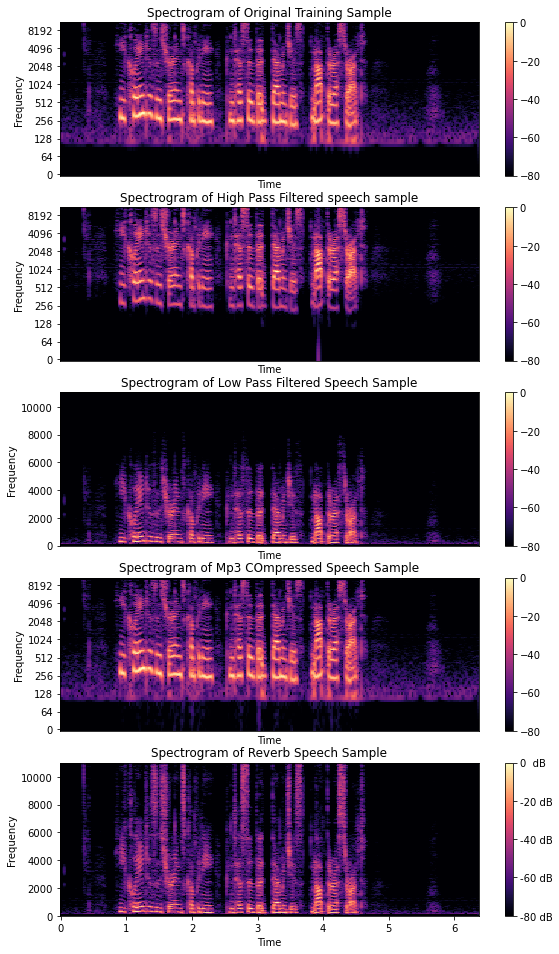}
        \caption{The spectrogramatic representation of each type of augmentation applied for the training of the network.}
        \label{fig:spectrogram}
\end{figure}

\subsection{Data augmentation}

 Data augmentation (DA) is a widely used technique in image and speech recognition to increase training data, prevent overfitting, and improve performance in class-imbalanced problems ~\cite{loshchilov2017decoupled}. To mitigate these challenges in our research, we employ five effective augmentation techniques: MP3 compression, high-pass filtering, low-pass filtering, silence trimming, and reverberation.

MP3 compression is known to be effective in spoofing detection~\cite{tak2022automatic}, and the other augmentations were selected based on their positive impact on model performance. High- and low-pass filtering, in particular, help extract sub-band information, which is crucial for detecting fine-grained features in state-of-the-art spoofing attacks. We also incorporate reverberation, aligning with the configuration settings used in the evaluation dataset preparation. The influence of each kind of the augmentation approach in spectral analysis is presented in Fig. \ref{fig:spectrogram}. The spectra of the speech samples demonstrates that the model was trained with diverse frequency sub-bands, which assists in learning the diverse artifacts of the speech samples. 
\subsection{Evaluation Metrics}
From the dataset details in Table~\ref{tab:Asvspoof2019}, it is clear that the ratio of real to spoofed trials is highly skewed. To address this, we employ alternative performance metrics, namely the Equal Error Rate (EER) and the Tandem Detection Cost Function (t-DCF), which are standard in ASVspoof challenges. In contrast to EER, t-DCF measures the performance evaluation of spoofing countermeasures (CMs) on the reliability of an ASV system. Wang et al.~\cite{wang2021comparative} demonstrate that the effectiveness of spoofing detection systems can vary significantly when random seeds are used. Similarly, after being trained with various random seeds, the EER of the baseline system in~\cite{tak2021end} fluctuates between $1.19\%$ and $2.06\%$. In response to these observations, the reported results in this study are an average of the best results obtained during the experiments with three random seeds.

\subsection{Experimental setup and hyper-parameters}
For all experiments, we utilize the Keras training platform in Python. Our anti-spoofing system employs the Adam optimizer with an initial learning rate of $1e^{-4}$ and a weight decay of $0.001$. The filter and kernel values for convolution layers are set to $64,128,256$ and $196,144,100$ for $c1,c2,c3$ and $k1,k2,k3$, respectively. We apply the cosine annealing warm restarts method~\cite{loshchilov2017decoupled} to adjust the learning rate, with linear growth for the first $1000$ warm-up steps, followed by a decrease according to the inverse square root of the step number. Our model is trained for $50$ epochs, using the cross-entropy loss function. Further, KAIMING initialization~\cite{he2015delving} is employed for all convolution layers, and batch normalization layers are configured with weights at $1$ and biases at $0$. The final model for evaluation is chosen based on the lowest loss observed on the development set.

We conducted all model training and testing on the Matilda High Performance Cluster at Oakland University. The HPC's GPU nodes, equipped with four NVIDIA Tesla V$100$ $16$G GPUs, $192$ GB of RAM, and $48$ $2.10$GHz CPU cores, were utilized for these tasks.

\section{Result and discussion}
In this section, we demonstrate the experimental and comparative results of our proposed anti-spoofing system. We optimized the hyper-parameters of our model as described above, and the results of the best set of hyper-parameters are provided below.\subsection{Trade-off between cardinalities and model width}
\begin{figure}[!t]
        \centering
        \includegraphics[width=8cm, height=5cm]{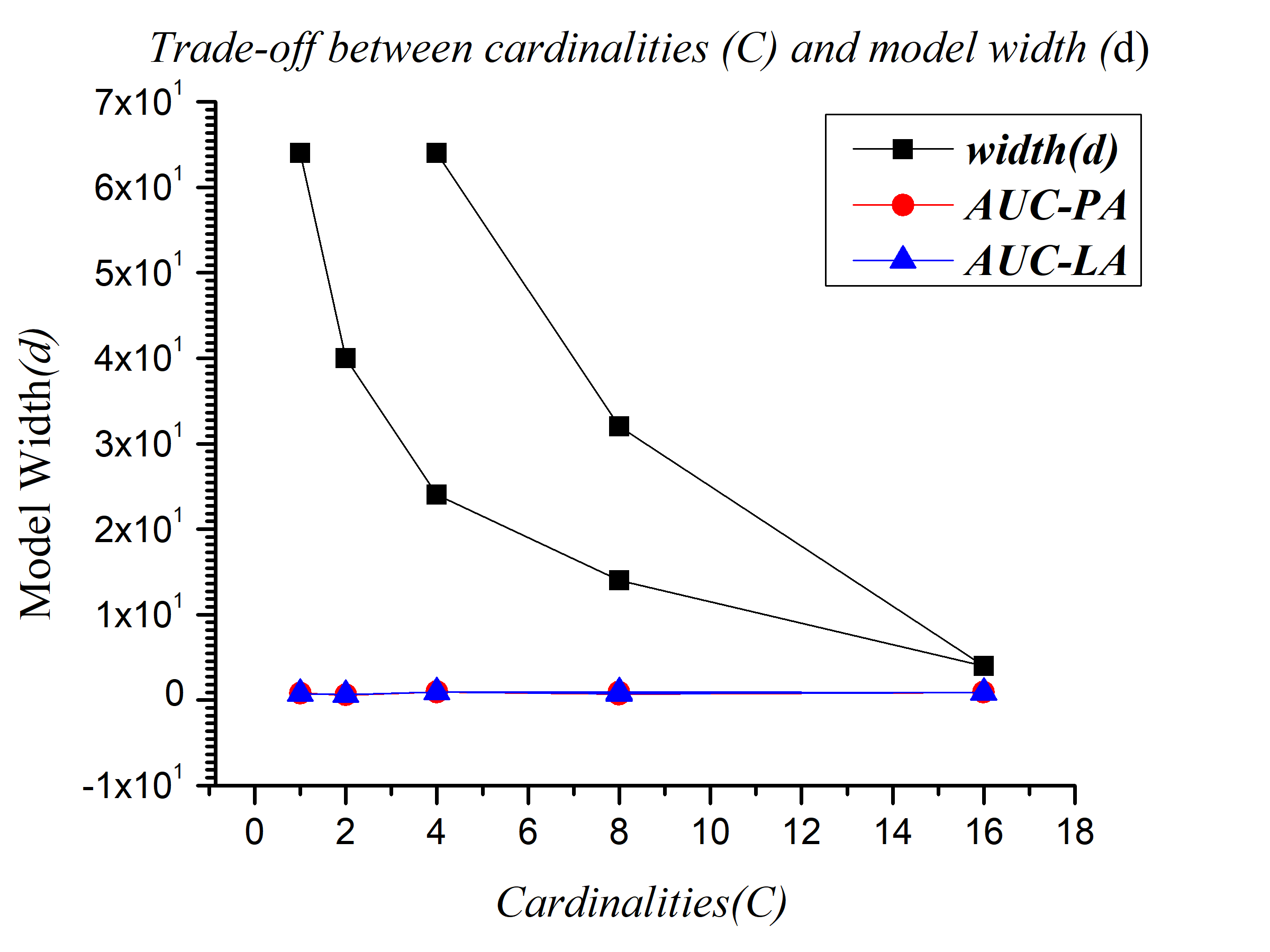}
        \caption{The model's performance is weighed against the trade-off between cardinality and model width. Before presenting, the model width is transformed to 1E3 ranges. The graph demonstrates that the AUC increases as the model width decreases.The model's performance is enhanced by lowering its complexity through a narrower model. }
        \label{fig:cardinality_graph}
\end{figure}

In split aggregate-based networks with multiple pathways, cardinality \textit{C} and bottleneck width \textit{d} are considered vital parameters. In contrast to the Inception network, which has unique cardinal paths, the proposed system is made up of pathways of varied cardinalities that follow the same configurations. Prior research has shown that split aggregate-based networks are more efficient with high cardinalities when evaluated against vision-based datasets like ImageNet and CIFAR. Therefore, as indicated in Table IV, we begin by examining the trade-off between cardinality and bottleneck width under conserved complexity.
\begin{table}[b]
\caption{Performance analysis of the proposed SE-PSA network with different cardinalities (\textit{C}) and model width (\textit{d}) to show the trade off in model effectiveness with ASVspoof2019 balanced speech samples.}
\label{tab:tradeoffcard}
\centering
\begin{tabular}{c |c | c | c}
  \hline
  Cardinality (\textit{C}) & Model-Width  (\textit{d})  & AUC-LA & AUC-PA \\
  \hline
 1             & 64 &  0.78 & 0.79 \\
 2              & 40 & 0.61 & 0.59  \\
 4              & 24   & 0.92 & 0.93 \\
 8              & 14  & 0.73 & 0.67\\
 16              & 4   & 0.87 & 0.89\\
 8              & 32  & 0.92 &  0.93\\
 \textbf{4}     & \textbf{64} &\textbf{ 0.93 }&  \textbf{0.97}\\
\hline      
\end{tabular}
 \end{table}
For this experiment, we used a balanced set of bonafide and spoofed speech samples from the LA and PA subsets of the ASVspoof2019 dataset. The results, reported in Table IV and  Fig. \ref{fig:cardinality_graph}, demonstrate that the $4\textit{C}\times64\textit{d}$ construction surpasses the other variants in terms of area under the curve (AUC) for spoofing detection. Specifically, this model obtained an AUC of 0.93\% and 0.97\% for the LA and PA subsets, respectively. Further, the results indicate that the AUC of the system fluctuates as the value of \textit{C} rises from 1 to 32. It is notable that when the value of \textit{C} was set to 1, the proposed network became equivalent to ResNet, as explained in Section 2. For the values of \textit{C} and \textit{d}, we chose the values proven to be effective against large-scale datasets, i.e., ImageNet and CIFAR \cite{xie2017aggregated}. The results showed that the $8\textit{C}\times32\textit{d}$ and $4\textit{C}\times24\textit{d}$ structures produced the second-best comparable results. Although these structures produced equivalent results for the LA spoofing samples, AUC degraded when detecting PA spoofing. 
\par Table IV and Fig. \ref{fig:cardinality_graph} further demonstrate that when the bottleneck width is small, increasing cardinality at the expense of decreasing width starts to produce saturated AUCs. Thus, from this result analysis, we conclude that raising the cardinalities and widths to higher values (as seen in \cite{xie2017aggregated}) is not worthwhile in the case of the ASVspoof2019 dataset.Optimum results are obtained when the cardinality ranges between 4 and 8 and the width is between 32 and 64.Consequently, the best-performing model has a cardinality of 4 and a model width of 64. In the next subsection, we present a performance analysis of the proposed system against familiar and unfamiliar attacks and the comparative performance.

\subsection{Performance analysis of the proposed anti-spoofing system}
\subsubsection{Performance analysis against Familiar spoofing Attacks}
In this experiment, we test the proposed method on the ASVspoof2019 and VSDC datasets. We use the training subsets of the both datasets for training, development subsets for validation, and evaluation subsets to test the effectiveness of the system. The results, in Table \ref{tab:performance-proposed}, demonstrate that the proposed system performs optimally, with an EER of 3.04\% and minimal t-DCF of 0.087 for the LA dataset, and an EER of 1.26 and min t-DCF of 0.038 when tested against PA speech samples. In the instance of the VSDC dataset, the proposed system obtains an EER of 0.32 for the first-order replays and an ideal EER of 0.87 when the speech sample contains the artifacts of multi-order replays. The results show that the proposed system performs better when tested against replay and chained replay spoofing attacks. This demonstrates the system's effectiveness against device artifacts in playback voice samples. Although the proposed system performs marginally better in PA attacks, it still surpasses the SOTA unified solutions.
\begin{table}[t]
\caption{Performance analysis of the proposed SE-PSA network against known spoofing Attacks from the ASVspoof2019 and VSDC datasets}
\label{tab:performance-proposed}
\centering
\begin{tabular}{*{6}{c}}
 \hline
  \multirow{2}{2cm}{Model} & \multicolumn{2}{c}{Cloning} & \multicolumn{2}{c}{Replay} & Chained Replay  \\ 
                         & EER & t-DCF & EER &t-DCF & EER  \\
\hline
 ASVspoof 2019 & 3.04 & 0.087 & 1.26 & 0.038 & --  \\
 VSDC & -- & -- & 0.32 & -- & 0.87  \\
\hline      
\end{tabular}
 \end{table}
 \begin{figure}[!b]
        \centering
        \includegraphics[width=8cm, height=5cm]{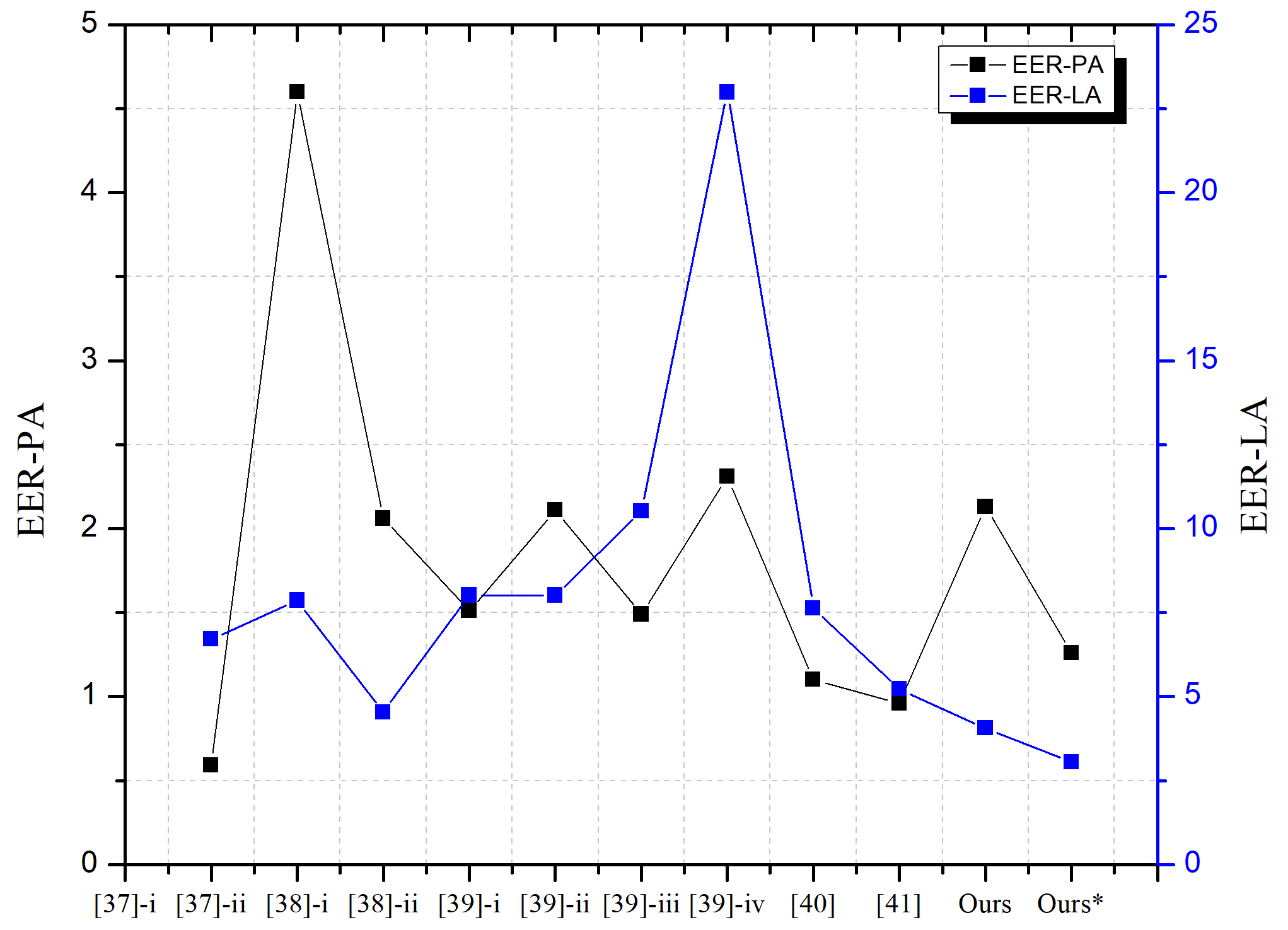}
        \caption{Comparative Analysis of the proposed PSA and state-of-the-art comparative methods, where * denotes augmentation. [37]-i and ii denote ASSERTS SENET and SENET-Resnet variation, [38]-i and ii denote FFT-CNN and LFCC-CNCC combination, and [39] i-iv denote logspec, SincNet, VGG, and SincNet with dropout, respectively.     }
        \label{fig:EERS2}
\end{figure}
\subsubsection{Performance analysis against unfamiliar spoofing attacks}
In this sub-experiment, we test the effectiveness of the proposed system in the absence of background knowledge about the spoofing attack. To the best of our knowledge, no unified model has been tested with integrated spoofing classes; SOTA systems report performance based on known clone and replay spoofing attacks.  Instead of training and testing the system separately, we combine LA and PA spoofing classes to create an integrated spoofing class. The model is trained to distinguish between bonafide, clone, and replay samples simultaneously within this integrated class. During testing, the proposed model confronts both LA and PA attacks together, resulting in an EER of $5.35\%$ and a t-DCF of $0.237$.

While the EER and minimum t-DCF are slightly higher compared to testing against known sample types (as mentioned earlier), this showcases the proposed model's applicability to real-world spoofing challenges. Furthermore, these results highlight the limitations of previous research, where separate training and testing fail to accurately assess the system's performance against various attack types. Specifically, when training and evaluation are restricted to either LA- or PA-based attacks, the performance of the model degrades significantly when evaluated against multiple spoofing attacks.
\begin{figure}[!t]
        \centering
        \includegraphics[width=8cm, height=4cm]{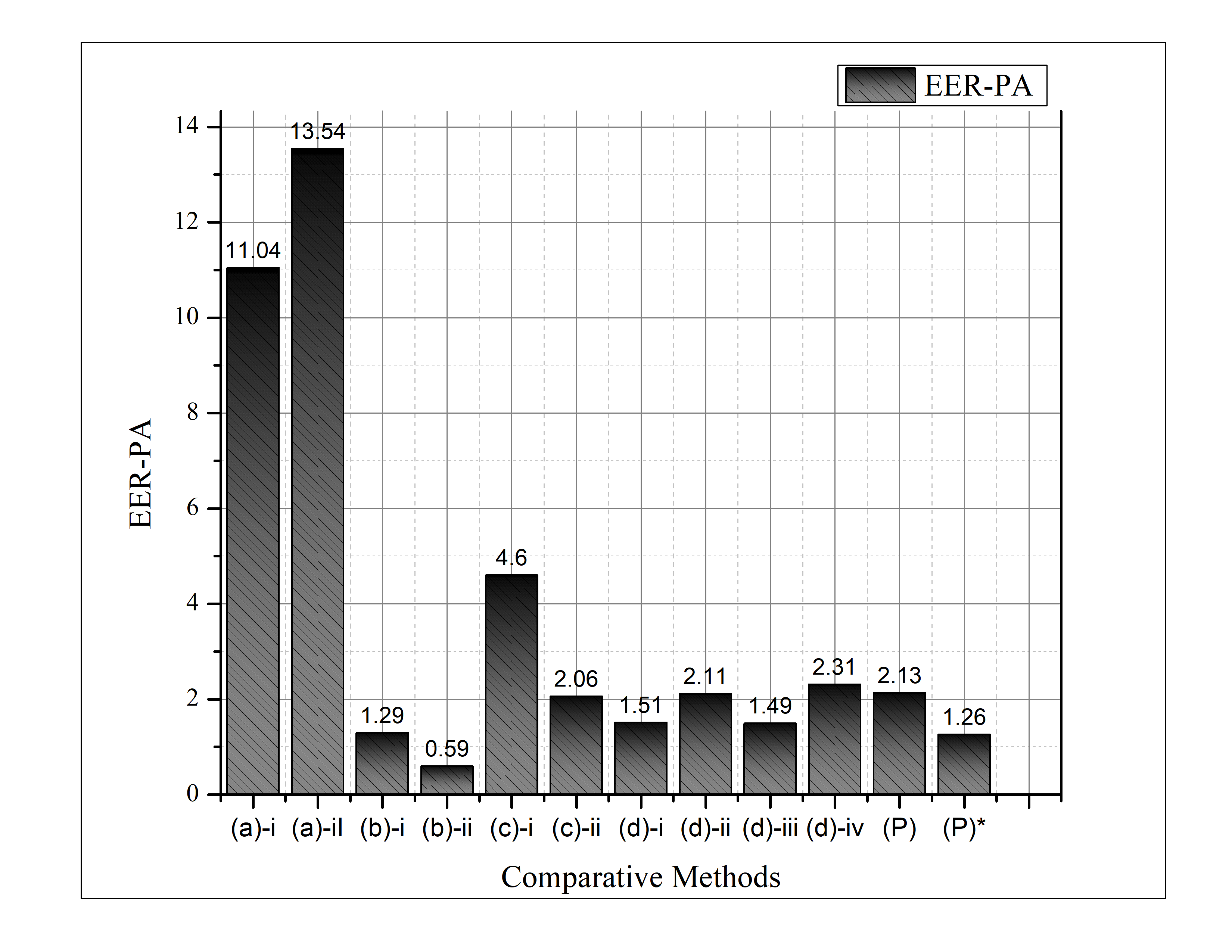}
        \caption{EER Comparative Analysis of SOTA systems against the Replay attack from ASVspoof2019,  where * denotes augmentation. (a)-i and ii denote Baseline ASVspoof2019 systems \cite{todisco2019asvspoof}, (b) i and ii are ASSERT variation \cite{lai2019assert}, (c)-i and ii denote STC \cite{lavrentyeva2019stc}, and (d) i-iv denote  \cite{zeinali2019detecting} logspec, SincNet, VGG, and SincNet with dropout, and (P) represents proposed systems, respectively.}
        \label{fig:EER_PA}
\end{figure}

\begin{figure}[!t]
        \centering
        \includegraphics[width=8cm, height=4cm]{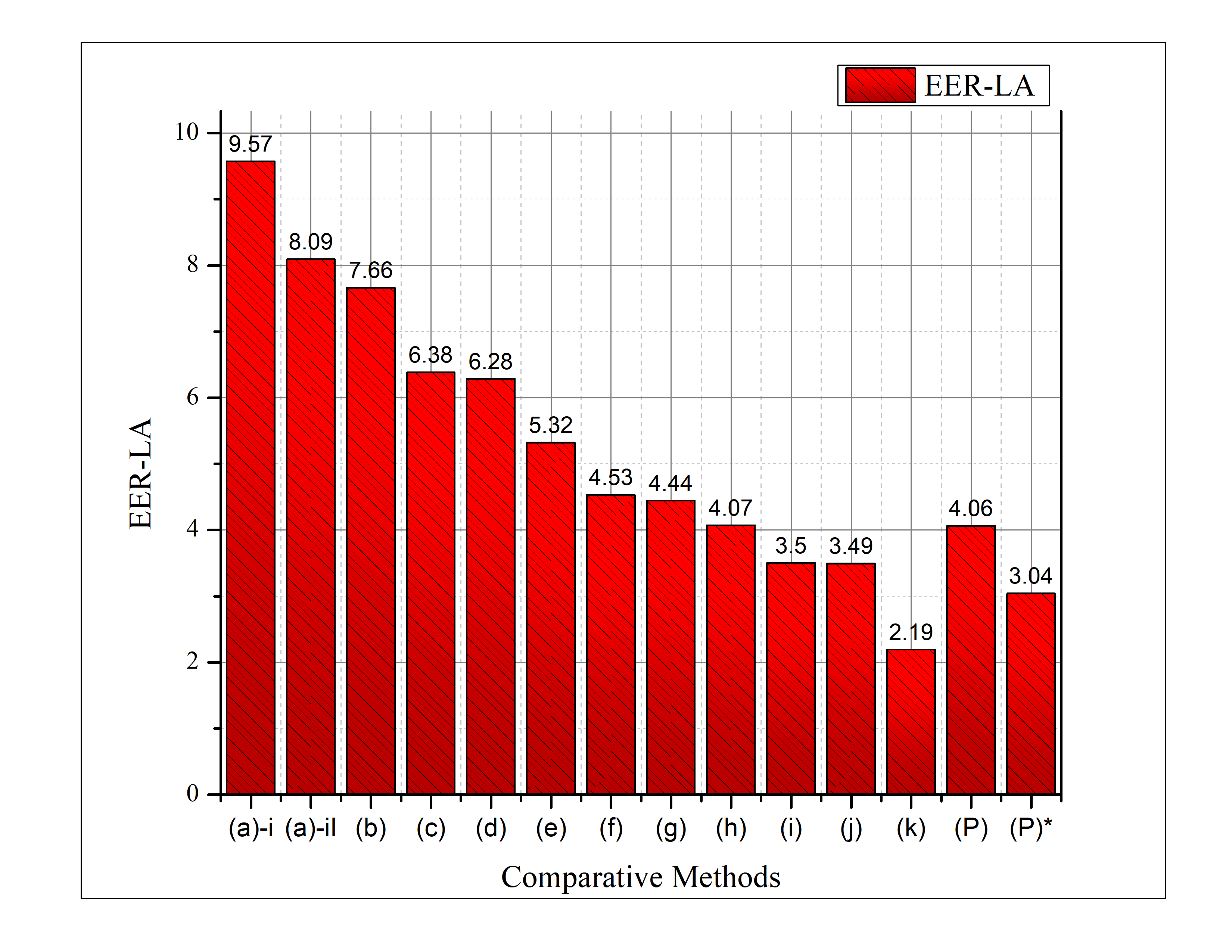}
        \caption{EER Comparative Analysis of SOTA systems against ASVspoof2019 synthesized and converted speech samples, where * denotes augmentation. (a)-i and ii denotes ASVspoof2019 baseline systems \cite{todisco2019asvspoof}, (b-k) denotes \cite{chettri2019ensemble}, \cite{monteiro2020generalized}, \cite{gomez2019light}, \cite{aravind2020audio},  \cite{lavrentyeva2019stc}, \cite{zhang2021one},    \cite{wu2020light}, \cite{tak2020spoofing}, \cite{chen2020generalization}, \cite{zhang2021one} and (P) represents proposed systems, respectively.  }
        \label{fig:EER_LA}
\end{figure}

\subsection{Performance analysis of the proposed and comparative methods against Logical Access (LA) spoofing attacks}
In this experiment, we examine the performance of the proposed model on synthetic and converted voice spoofing samples. The model is trained using the training subset of the ASVspoof-LA dataset, along with five types of augmented samples (as described in Section II). The proposed system is compared with twelve comparative methods, and the results are presented in Table \ref{tab:LAresults}. The results demonstrate that, when trained using augmented samples, the proposed system obtains an EER of 3.04\% and a t-DCF of 0.087\%, whereas without augmented samples, the model achieves 4.06\% and 0.099\%, respectively. These results indicate that the proposed system outperformed eleven of the twelve SOTA comparative countermeasures, with the lowest EER and t-DCF. More specifically, the proposed system performed second best on the ASVspoof2019 LA dataset, both with and without augmented samples. The EER and minimum t-DCF of comparative methods are reported in Table \ref{tab:LAresults} and Fig. \ref{fig:EER_LA}. Despite having a slightly lower EER and minimal t-DCF than the proposed system, \cite{zhang2021one} is particularly optimized to identify LA-based attacks and has never been evaluated against replay attacks. In contrast, the proposed system obtained a lower EER and minimal t-DCF even when trained without any type of augmentation. This indicates the proposed system's superiority as a robust countermeasure to voice cloning and conversion attacks. In the next section, we compare the performance of the proposed system against SOTA replay attack detection systems.

\begin{table}[t]
\caption{Performance comparison of the proposed PSA network with state-of-the-art comparative methods against ASVspoof2019 synthesized and converted speech samples, where * refers to training with augmented samples.}
\label{tab:LAresults}
\centering
\begin{tabular}{c |c | c | c}
  \hline
  Method & Input   & EER & min T-DCF \\
  \hline
 Baseline-Asvspoof2019 \cite{todisco2019asvspoof} & CQCC+GMM & 9.57 & 0.237  \\
 Baseline-Asvspoof2019 \cite{todisco2019asvspoof}& LFCC+GMM  & 8.09 & 0.212 \\
 Chettri et al. \cite{chettri2019ensemble} & Spatial features & 7.66 & 0.179\\
 Monterio et al. \cite{monteiro2020generalized} & Spectrogram  & 6.38 & 0.142\\
 Gomez-Alanis et al.\cite{gomez2019light} & Spectrogram  & 6.28 &  -\\
 Aravind et al.    \cite{aravind2020audio} & Mel-spectrogram & 5.32 & 0.151\\
 Lavrentyeva et al.  \cite{lavrentyeva2019stc}   & CQT & 4.53 & 0.103\\
 ResNet + OC-SVM  \cite{zhang2021one}  & -- & 4.44 & 0.115\\
 Wu et al  \cite{wu2020light}  & deep features & 4.07 & 0.102\\
 Tak et al.  \cite{tak2020spoofing}   & LFCC & 3.50 & 0.090\\
 Chen et al.  \cite{chen2020generalization}  & Filter banks & 3.49 & 0.092\\
 One class Learning \cite{zhang2021one}& LFCC-60D &  2.19 & 0.059 \\
\textbf{ PSA-18 layers }   & Raw audio & 4.06 & 0.099\\
\textbf{PSA-18 layers*}&\textbf{Raw audio} & \textbf{3.04 }& \textbf{0.087} \\
 \hline      
\end{tabular}
 \end{table}
 
\subsection{Performance analysis of the proposed and comparative methods against Physical Access (PA) spoofing attacks}
In this experiment, we test the proposed PSA system's resilience against replay spoofing attacks. The proposed system is trained using the training subset of the ASVspoof-PA dataset, validated using the development subset, and tested using the evaluation subset of the dataset. The results, shown in Table \ref{tab:PAresults}, indicate that the proposed system effectively discriminates between spoofed and bonafide artifacts in replayed voice samples. When trained using augmented samples, the proposed system achieves an optimal EER of 1.26\% and a minimal t-DCF of 0.038\%. In comparison, the proposed system attains an EER of 2.13\% and a t-DCF of 0.064\% without augmentation. These results show that the proposed system achieved the second-lowest EER, after ASSERT [37]. Although the EER of the ASSERT solution is slightly better in PA spoofing attacks, the proposed model outperformed ASSERT in LA spoofing attacks. Further, the ASSERT model is based on handmade features and a 50-layer SENet architecture, whereas the proposed model has an 18-layer architecture and can extract the required deep features directly from raw audio. Except for the ASSERT, the proposed model outperformed the other eight comparative models in replay speech detection. The results in terms of EER and t-DCF of other comparative methods are described in Table \ref{tab:PAresults} and Fig. \ref{fig:EER_PA}. 

In the next subsection, we compare the proposed system's performance to SOTA unified methods designed to identify both LA and PA spoofing attacks.
\begin{figure}[!b]
        \centering
        \includegraphics[width=8cm, height=5cm]{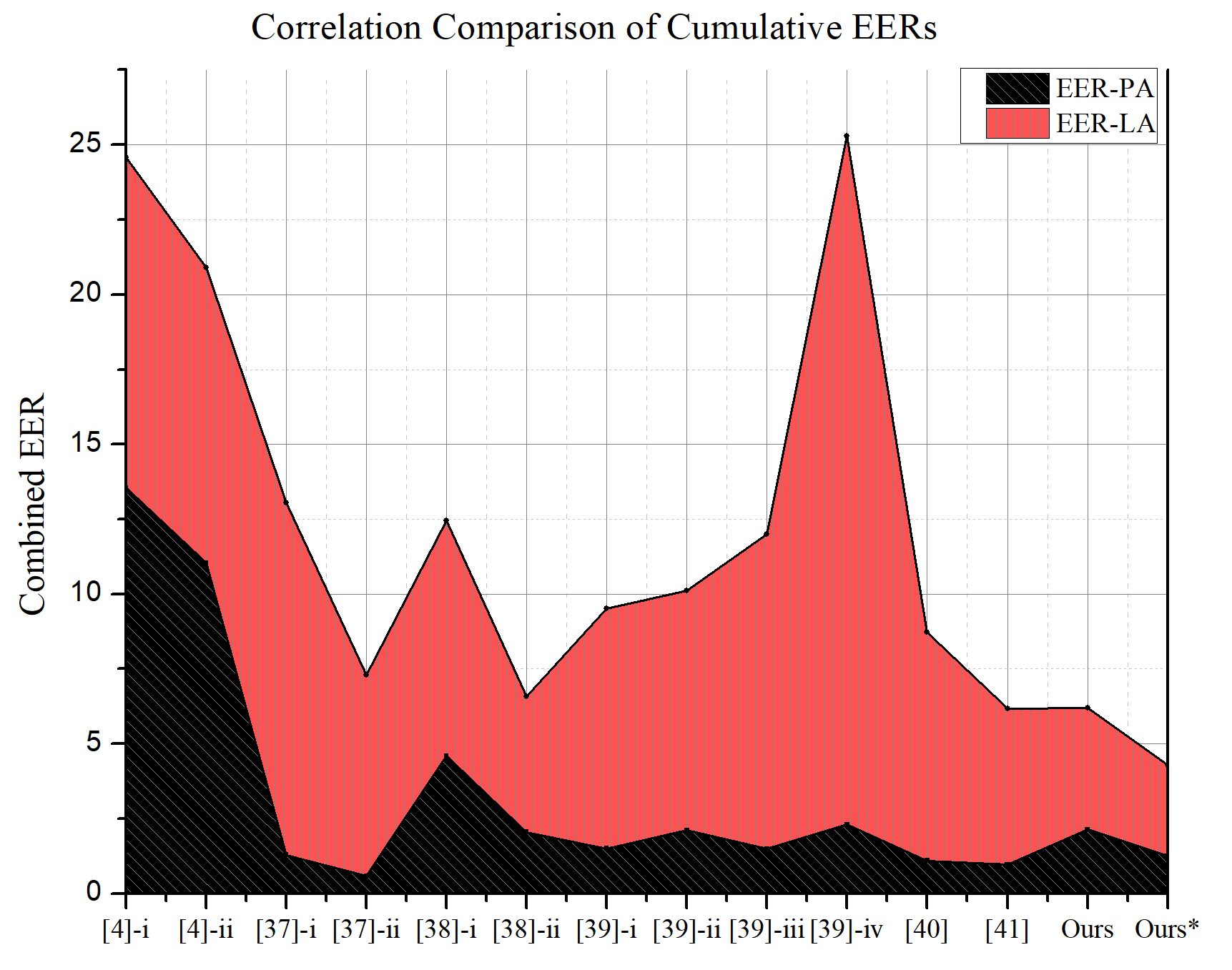}
        \caption{Comparison of the cumulative EERs of SOTA's unified anti-spoofing systems. The graph illustrates that, when compared to other approaches, the proposed method has the lowest cumulative EER, where * denotes augmentation. [37]-i and ii denote ASSERTS SENET and SENET-Resnet variation, [38]-i and ii denote FFT-CNN and LFCC-CNCC combination, and [39] i-iv denote logspec, SincNet, VGG, and SincNet with dropout, respectively.   }
        \label{fig:commulativeEER}
\end{figure}
\begin{table}[t]
\caption{Performance comparison of the proposed PSA network with state-of-the-art (SOTA) comparative methods against Replay Attacks; * refers to training with augmented samples.}
\label{tab:PAresults}
\centering
\begin{tabular}{p{2.5cm} |p{2.8cm}  | p{0.6cm}  | p{1.0cm} }
  \hline
  Method & Input   & EER & min T-DCF \\
  \hline
 Baseline-Asvspoof2019 \cite{todisco2019asvspoof} & CQCC+GMM & 11.04 & 0.2454  \\
 Baseline-Asvspoof2019 \cite{todisco2019asvspoof}& LFCC+GMM  & 13.54 & 0.3017 \\
 ASSERT \cite{lai2019assert} & log-spec-SENet & 1.29 & 0.036\\
 ASSERT \cite{lai2019assert} & SENet50-Dialated ResNet  & 0.59 & 0.016\\
 STC  \cite{lavrentyeva2019stc}   & LFCC-CMVN-LCNN & 4.6 & 0.105\\
 STC  \cite{lavrentyeva2019stc}   & FFT-LCNN & 2.06 & 0.56\\
 BUT-Omilia  \cite{zeinali2019detecting}  & logSpec-VGG-SincNet 1
-SincNet 2 & 1.51 & 0.0372\\
 BUT-Omilia  \cite{zeinali2019detecting}  & SincNet with standard
dropout & 2.11 & 0.052\\
 BUT-Omilia  \cite{zeinali2019detecting}  & VGG 1-VGG 2 & 1.49 & 0.04\\
 BUT-Omilia  \cite{zeinali2019detecting}  & SincNet with high dropout & 2.31 & 0.059\\
\textbf{PSA-18 layers  }   & Raw audio & 2.13 & 0.064\\
\textbf{PSA-18 layers*}&\textbf{Raw audio} & \textbf{1.26}& \textbf{0.038} \\
 \hline      
\end{tabular}
 \end{table}

\subsection{Correlation and Cumulative performance analysis of proposed and comparative unified solutions}
In this experiment, we evaluate the performance of the proposed system compared to seven state-of-the-art unified solutions designed for detecting both LA and PA spoofing attacks. The results, summarized in Table~\ref{tab:unifiedresults}, clearly demonstrate the superiority of the proposed method in terms of EER and minimal t-DCF metrics.

When we use data augmentation during training, the proposed system achieves the lowest EER and t-DCF values for synthetic and converted speech samples, at $3.04\%$ and $0.087\%$, respectively. Without augmentation, the proposed system still maintains robust performance, with EER and t-DCF values of $4.06\%$ and $0.99\%$, respectively, especially against voice cloning attacks.

In the case of PA spoofing attacks, the proposed system again outperforms the competition with an EER of $1.26\%$ and a t-DCF of $0.038\%$ when trained with augmentation, and EER and t-DCF values of $2.13\%$ and $0.064\%$, respectively, without augmentation. Additionally, we evaluated the cumulative EERs (combined EERs for PA and LA attacks) of the existing solutions, highlighting that the proposed system achieves an impressive EER of nearly $4.30\%$, surpassing the SOTA unified methods. This is depicted in Fig. cumulative EER.

\begin{figure}[!b]
        \centering
        \includegraphics[width=8cm, height= 5cm]{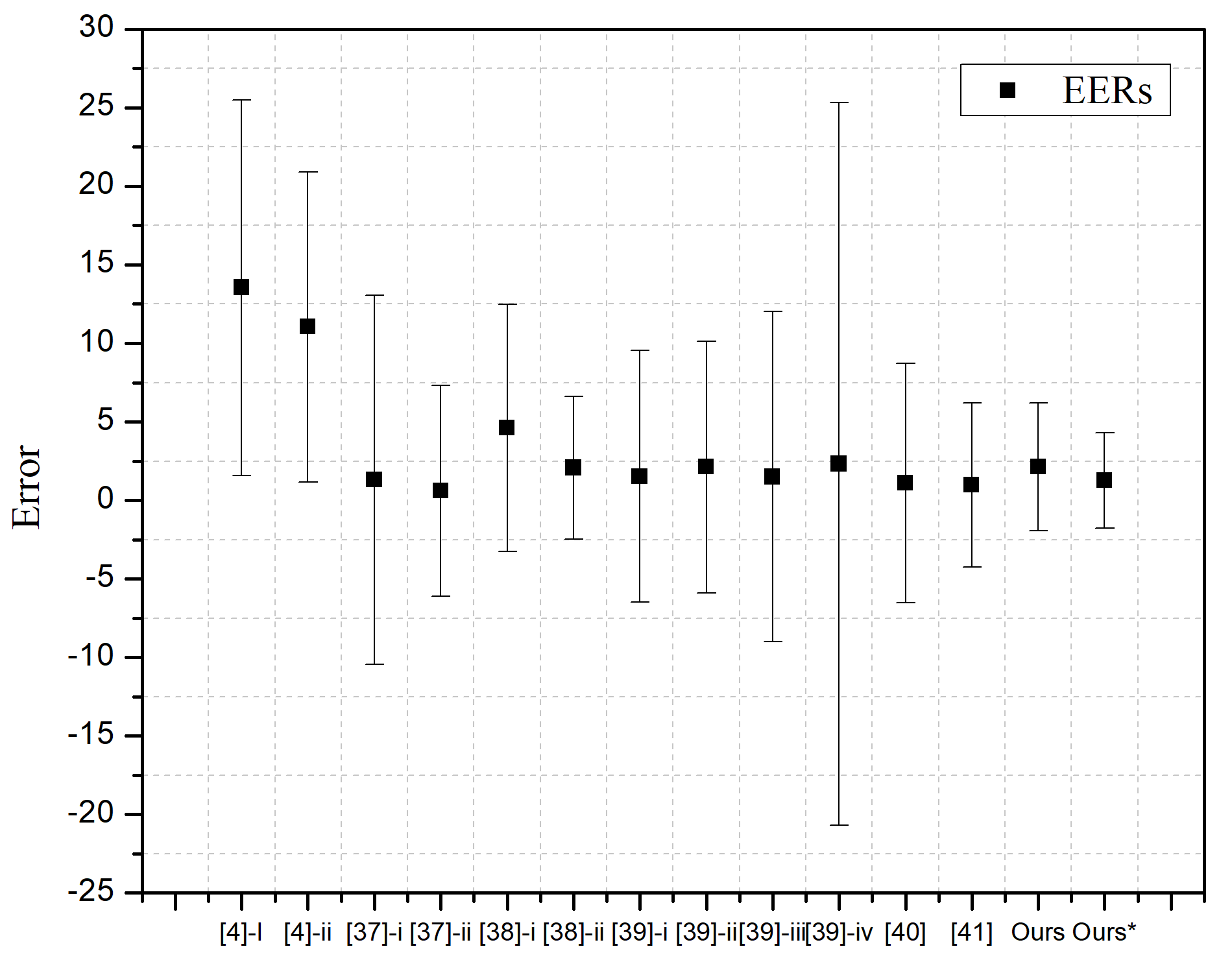}
        \caption{Error bar graph representation of proposed and SOTA's unified solutions, where * denotes augmentation. [37]-i and ii denote ASSERTS SENET and SENET-Resnet variation, [38]-i and ii denote FFT-CNN and LFCC-CNCC combination, and [39] i-iv denote logspec, SincNet, VGG, and SincNet with dropout, respectively.}
        \label{fig:ERRofEER}
\end{figure}

To provide a comprehensive view of overall performance, we created an error bar graph for all unified solutions, once again showing the superiority of the proposed solution in detecting both LA and PA attacks, as seen in Fig. \ref{fig:ERRofEER}. Comparisons with other unified approaches reveal that many of them exhibit a significant EER disparity of over $4\%$ between LA and PA spoof sample detection, except for STC \cite{lavrentyeva2019stc}, which has a $2\%$ EER deviation. The proposed system, on the other hand, effectively reduces the EER disparities between LA and PA attacks, achieving the best detection results. While the EER of the proposed system is slightly higher than that of SASV \cite{aljasem2021secure}, ASSERT \cite{lai2019assert}, and MFMT \cite{li2019anti}, it significantly outperforms these systems in LA attack detection. It's worth noting that none of these systems were designed as end-to-end solutions; they all rely on computationally expensive handcrafted feature extraction. This highlights the proposed system's superiority over the current state-of-the-art unified countermeasures against LA and PA attacks.
\begin{table}[t]
\caption{Experimental performance of the proposed system against state-of-the-art unified voice spoofing countermeasures.}
\label{tab:unifiedresults}
\centering
\begin{tabular}{p{2.2cm}|p{1.0cm}|p{1.3cm}|p{1.0cm}|p{1.3cm}} %{*{6}{l}}
\hline
 \multirow{2}{1cm}{Paper}  & \multicolumn{2}{c}{Logical Acces} & \multicolumn{2}{c}{Physical Access} \\ 
                         & EER & min T-DCF & EER & min T-DCF  \\
 \hline
 Baseline \cite{todisco2019asvspoof}& 11.96 & 0.212 & 13.54  & 0.3017    \\
 Baseline \cite{todisco2019asvspoof} & 9.87 & 0.236 & 11.04  & 0.2454   \\
 ASSERT \cite{lai2019assert}  & 11.75 & 0.216 & 1.29  & 0.036  \\
 ASSERT \cite{lai2019assert}  & 6.70 & 0.155 & 0.59  & 0.016  \\
 STC  \cite{lavrentyeva2019stc}   & 7.86  & 0.183 & 4.6  & 0.105  \\
 STC  \cite{lavrentyeva2019stc}  & 4.53 & 0.103 & 2.06 & 0.56 \\
 BUT-Omilia  \cite{zeinali2019detecting} & 8.01 &0.208& 1.51& 0.0372  \\
 BUT-Omilia  \cite{zeinali2019detecting}  & 8.01 &0.356 &2.11& 0.0527 \\
 BUT-Omilia  \cite{zeinali2019detecting}  & 10.52 &0.279& 1.49& 0.04 \\
 BUT-Omilia  \cite{zeinali2019detecting}  & 22.99& 0.381& 2.31 &0.0591 \\
MFMT \cite{li2019anti}   & 7.63 &0.213 &0.96 &0.0266 \\
 SASV  \cite{aljasem2021secure} & 5.22& 0.132 & 1.1 &0.0335 \\
 \textbf{PSA-18 layers}     & 4.06 & 0.099&  2.13 & 0.064  \\
\textbf{PSA-18 layers*}& \textbf{3.04 }& \textbf{0.087} & \textbf{1.26}& \textbf{0.038} \\
  \hline
\end{tabular}

\end{table}

\subsection{Performance comparison of PSA with Hand Crafted features}
In this experiment, we feed the traditional handcrafted features into the proposed SE-PSA network to see the effectiveness of the system. These handcrafted features have proven to be successful in voice spoof detection; however, their efficacy within the aggregated network has yet to be investigated. In order to demonstrate the efficiency of the PSA network against handcrafted features, we examined 5 handcrafted features: CQCC, LFCC, MFCC, GTCC, and LPCC. A 20-number filter is used for all features except for CQCC, where a 96-octave filter is used. We use mean aggregation to transform the 2D feature specifications into 1D before feeding them into the network. We examine the performance of the PSA network with a balanced subset of the ASVspoof2019-LA dataset, and the results are shown in Table \ref{tab:handcraft}. These results indicate that the aggregated network incorporates the higher-order distinctions of the feature map fed into it. The local and global transformations of the feature map need a larger input to be performed optimally. As a result, the raw wave input with the appropriate number of frame lengths performs better than the other handcrafted features. The results show that the LFCC and GTCC features perform well, with an EER of 10.09 and 7.66, respectively; however, this EER is higher than raw wave audio. In contrast, the MFCC features show the highest EER, 26.38, and a minimum t-DCF of 0.352. Thus, we can conclude that the designed aggregated network performs better with raw waveforms compared to handcrafted features.

 \begin{table}
 \caption{Experimental performance comparison of the PSA countermeasure with raw audio and traditional handcrafted features.}
\label{tab:handcraft}
\centering
\begin{tabular}{c | c | c}
  \hline
  Input   & EER & min T-DCF \\
  \hline
 CQCC & 11.57 & 0.317  \\
 LFCC  & 10.09 & 0.292 \\
GTCC & 7.66 & 0.179\\
 MFCC  & 26.38 & 0.352\\
 LPCC  & 16.38 & 0.242\\
 \textbf{Raw audio} & \textbf{2.65} & \textbf{0.079}\\
 \hline      
\end{tabular}
 \end{table}

\subsection{Ablation Study}
Different channel cardinality combinations, dropouts, and layer topologies were investigated to avoid over-fitting and under-fitting during training. To achieve the intended goals, we evaluated the SE-PSA design by increasing the width and density of the network. In all, we evaluated the resnet architectures with 18, 34, 50, and 101 layers as well as the aggregated network with identical layered structures. All of the networks were trained for 20 epochs using the ASVspoof-LA datasets. The results showed that training the larger network required a significantly larger number of training epochs and more data, and the models became overfit after training with the speech samples available in the LA subset. However, the aggregated network with SE and skip connections, with 18 and 34 layers of architecture, respectively, outperformed state-of-the-art networks. In comparison to all other approaches, the aggregated network with 34 layers performs the second best with an EER of 8.54\% while the lowest EER is obtained by the SE-aggregated network with spatial dropout. Surprisingly, when fed raw audio samples, the ResNet architecture failed to perform properly and obtained a higher EER.

We evaluated networks with and without SE and skip connections in addition to varying network densities, as shown in Table \ref{tab:ablationstudy}. The aggregated network consistently outperformed ResNet networks, with notable EER differences (detailed in Table \ref{tab:ablationstudy}). While ResNets with SE connections showed improved performance compared to those without, the aggregated network consistently achieved better EER results. For instance, SE-Resnet variants achieved EERs of 6.87, 12.66, 23.54, and 30.33, while the aggregated networks achieved 5.50, 6.43, 29.65, and 29.54, respectively.

To prevent overfitting due to numerous cardinalities and aggregation, we used various strategies, including spatial dropout after the aggregation layer, resulting in superior results with an EER of 4.06 when combined with SE and skip connections.
\begin{table}[t]
\caption{ResNets and ResNext models with 18, 34, 50, and 101 layers testing with raw audio, SE and SKIP connections, and the impact of spatial dropout during aggregation testing}
\label{tab:ablationstudy}
\centering
\begin{tabular}{c |c  }
  \hline
  Cardinality & EER  \\ \hline
ResNet-18            & 11.04    \\
SE-ResNet-18             & 6.87   \\
ResNet-34       & 17.54   \\
SE-ResNet-34        & 12.66  \\
ResNet-50         & 38.43  \\
SE-ResNet-50       & 23.54 \\
ResNet-101       & 38.66   \\
SE-ResNet-101        & 30.33 \\
Aggregated Nets-18            & 7.65  \\
SE-Aggregated Nets-18             & 5.50  \\
Aggregated Nets-34       & 8.54  \\
SE-Aggregated Nets-34        & 6.43\\
Aggregated Nets-50         & 30.75 \\
SE-Aggregated Nets-50       & 29.65 \\
Aggregated Nets-101       & 32.76  \\
SE-Aggregated Nets-101        & 29.54 \\
Aggregated Nets-18 (Spatial Dropout)  & 6.4 \\
\textbf{SE-Aggregated Nets-18 (Spatial Dropout)} & \textbf{4.06}   \\
\hline      
\end{tabular}
 \end{table}

\section{Conclusion}
This paper introduces a unified spoofing detection system, using a Parallel Stack Aggregation (PSA) network to process raw audio directly. The method employs a Split-Transform-Merge (STM) strategy with multiple cardinal points to effectively learn logical and physical artifacts from speech samples. Experimental results on the ASVspoof 2019 and VSDC datasets demonstrate that the proposed anti-spoofing model significantly outperforms both baselines and state-of-the-art systems. Additionally, the proposed network reduces the intra-EER distinction between logical and physical attacks, detecting both equally effectively. Future work aims to extend the system to include liveliness detection and automatic speaker verification.
 \section{Acknowledgement}
This study is funded by NSF award number 1815724 and MTRAC ACT award number 292883. The opinions, results, conclusions, or recommendations in this material are solely those of the author(s) and do not necessarily represent NSF or MTRAC ACT views.
\bibliographystyle{IEEEtran}
\bibliography{references.bib}

\begin{thebibliography}{10}
\providecommand{\url}[1]{#1}
\csname url@samestyle\endcsname
\providecommand{\newblock}{\relax}
\providecommand{\bibinfo}[2]{#2}
\providecommand{\BIBentrySTDinterwordspacing}{\spaceskip=0pt\relax}
\providecommand{\BIBentryALTinterwordstretchfactor}{4}
\providecommand{\BIBentryALTinterwordspacing}{\spaceskip=\fontdimen2\font plus
\BIBentryALTinterwordstretchfactor\fontdimen3\font minus
  \fontdimen4\font\relax}
\providecommand{\BIBforeignlanguage}[2]{{%
\expandafter\ifx\csname l@#1\endcsname\relax
\typeout{** WARNING: IEEEtran.bst: No hyphenation pattern has been}%
\typeout{** loaded for the language `#1'. Using the pattern for}%
\typeout{** the default language instead.}%
\else
\language=\csname l@#1\endcsname
\fi
#2}}
\providecommand{\BIBdecl}{\relax}
\BIBdecl

\bibitem{obaidat2019biometric}
M.~S. Obaidat, S.~P. Rana, T.~Maitra, D.~Giri, and S.~Dutta, ``Biometric
  security and internet of things (iot),'' in \emph{Biometric-based physical
  and cybersecurity systems}.\hskip 1em plus 0.5em minus 0.4em\relax Springer,
  2019, pp. 477--509.

\bibitem{sahidullah2019introduction}
M.~Sahidullah, H.~Delgado, M.~Todisco, T.~Kinnunen, N.~Evans, J.~Yamagishi, and
  K.-A. Lee, ``Introduction to voice presentation attack detection and recent
  advances,'' \emph{Handbook of biometric anti-spoofing}, pp. 321--361, 2019.

\bibitem{khan2023battling}
A.~Khan, K.~M. Malik, J.~Ryan, and M.~Saravanan, ``Battling voice spoofing: a
  review, comparative analysis, and generalizability evaluation of
  state-of-the-art voice spoofing counter measures,'' \emph{Artificial
  Intelligence Review}, pp. 1--54, 2023.

\bibitem{das2020assessing}
R.~K. Das, J.~Yang, and H.~Li, ``Assessing the scope of generalized
  countermeasures for anti-spoofing,'' in \emph{ICASSP 2020-2020 IEEE
  International Conference on Acoustics, Speech and Signal Processing
  (ICASSP)}.\hskip 1em plus 0.5em minus 0.4em\relax IEEE, 2020, pp. 6589--6593.

\bibitem{witkowski2017audio}
M.~Witkowski, S.~Kacprzak, P.~Zelasko, K.~Kowalczyk, and J.~Galka, ``Audio
  replay attack detection using high-frequency features.'' in
  \emph{Interspeech}, 2017, pp. 27--31.

\bibitem{zhang122021effect}
Y.~Zhang12, W.~Wang12, and P.~Zhang12, ``The effect of silence and dual-band
  fusion in anti-spoofing system,'' 2021.

\bibitem{jung2022aasist}
J.-w. Jung, H.-S. Heo, H.~Tak, H.-j. Shim, J.~S. Chung, B.-J. Lee, H.-J. Yu,
  and N.~Evans, ``Aasist: Audio anti-spoofing using integrated spectro-temporal
  graph attention networks,'' in \emph{ICASSP 2022-2022 IEEE International
  Conference on Acoustics, Speech and Signal Processing (ICASSP)}.\hskip 1em
  plus 0.5em minus 0.4em\relax IEEE, 2022, pp. 6367--6371.

\bibitem{tak2021end}
H.~Tak, J.-w. Jung, J.~Patino, M.~Kamble, M.~Todisco, and N.~Evans,
  ``End-to-end spectro-temporal graph attention networks for speaker
  verification anti-spoofing and speech deepfake detection,'' \emph{arXiv
  preprint arXiv:2107.12710}, 2021.

\bibitem{8272715}
H.~Muckenhirn, M.~Magimai-Doss, and S.~Marcel, ``End-to-end convolutional
  neural network-based voice presentation attack detection,'' in \emph{2017
  IEEE International Joint Conference on Biometrics (IJCB)}, 2017, pp.
  335--341.

\bibitem{lai2019assert}
C.-I. Lai, N.~Chen, J.~Villalba, and N.~Dehak, ``Assert: Anti-spoofing with
  squeeze-excitation and residual networks,'' \emph{arXiv preprint
  arXiv:1904.01120}, 2019.

\bibitem{lavrentyeva2019stc}
G.~Lavrentyeva, S.~Novoselov, A.~Tseren, M.~Volkova, A.~Gorlanov, and
  A.~Kozlov, ``Stc antispoofing systems for the asvspoof2019 challenge,''
  \emph{arXiv preprint arXiv:1904.05576}, 2019.

\bibitem{zeinali2019detecting}
H.~Zeinali, T.~Stafylakis, G.~Athanasopoulou, J.~Rohdin, I.~Gkinis, L.~Burget,
  J.~{\v{C}}ernock{\`y} \emph{et~al.}, ``Detecting spoofing attacks using vgg
  and sincnet: but-omilia submission to asvspoof 2019 challenge,'' \emph{arXiv
  preprint arXiv:1907.12908}, 2019.

\bibitem{li2019anti}
R.~Li, M.~Zhao, Z.~Li, L.~Li, and Q.~Hong, ``Anti-spoofing speaker verification
  system with multi-feature integration and multi-task learning.'' in
  \emph{Interspeech}, 2019, pp. 1048--1052.

\bibitem{aljasem2021secure}
M.~Aljasem, A.~Irtaza, H.~Malik, N.~Saba, A.~Javed, K.~M. Malik, and
  M.~Meharmohammadi, ``Secure automatic speaker verification (sasv) system
  through sm-altp features and asymmetric bagging,'' \emph{IEEE Transactions on
  Information Forensics and Security}, vol.~16, pp. 3524--3537, 2021.

\bibitem{xie2017aggregated}
S.~Xie, R.~Girshick, P.~Doll{\'a}r, Z.~Tu, and K.~He, ``Aggregated residual
  transformations for deep neural networks,'' in \emph{Proceedings of the IEEE
  conference on computer vision and pattern recognition}, 2017, pp. 1492--1500.

\bibitem{khan2022voice}
A.~Khan, K.~M. Malik, J.~Ryan, and M.~Saravanan, ``Voice spoofing
  countermeasures: Taxonomy, state-of-the-art, experimental analysis of
  generalizability, open challenges, and the way forward,'' \emph{arXiv
  preprint arXiv:2210.00417}, 2022.

\bibitem{li2021replay}
X.~Li, N.~Li, C.~Weng, X.~Liu, D.~Su, D.~Yu, and H.~Meng, ``Replay and
  synthetic speech detection with res2net architecture,'' in \emph{ICASSP
  2021-2021 IEEE international conference on acoustics, speech and signal
  processing (ICASSP)}.\hskip 1em plus 0.5em minus 0.4em\relax IEEE, 2021, pp.
  6354--6358.

\bibitem{wu2012detecting}
Z.~Wu, E.~S. Chng, and H.~Li, ``Detecting converted speech and natural speech
  for anti-spoofing attack in speaker recognition,'' in \emph{Thirteenth Annual
  Conference of the International Speech Communication Association}, 2012.

\bibitem{sriskandaraja2016front}
K.~Sriskandaraja, V.~Sethu, E.~Ambikairajah, and H.~Li, ``Front-end for
  antispoofing countermeasures in speaker verification: Scattering spectral
  decomposition,'' \emph{IEEE Journal of Selected Topics in Signal Processing},
  vol.~11, no.~4, pp. 632--643, 2016.

\bibitem{yang2021modified}
J.~Yang, H.~Wang, R.~K. Das, and Y.~Qian, ``Modified magnitude-phase spectrum
  information for spoofing detection,'' \emph{IEEE/ACM Transactions on Audio,
  Speech, and Language Processing}, vol.~29, pp. 1065--1078, 2021.

\bibitem{kamble2020amplitude}
M.~R. Kamble, H.~Tak, and H.~A. Patil, ``Amplitude and frequency
  modulation-based features for detection of replay spoof speech,''
  \emph{Speech Communication}, vol. 125, pp. 114--127, 2020.

\bibitem{sahidullah2015comparison}
M.~Sahidullah, T.~Kinnunen, and C.~Hanil{\c{c}}i, ``A comparison of features
  for synthetic speech detection,'' 2015.

\bibitem{khan2022toward}
A.~Khan, A.~Javed, K.~M. Malik, M.~A. Raza, J.~Ryan, A.~K.~J. Saudagar, and
  H.~Malik, ``Toward realigning automatic speaker verification in the era of
  covid-19,'' \emph{Sensors}, vol.~22, no.~7, p. 2638, 2022.

\bibitem{rahmeni2020speech}
R.~Rahmeni, A.~B. Aicha, and Y.~B. Ayed, ``Speech spoofing detection using svm
  and elm technique with acoustic features,'' in \emph{2020 5th International
  Conference on Advanced Technologies for Signal and Image Processing
  (ATSIP)}.\hskip 1em plus 0.5em minus 0.4em\relax IEEE, 2020, pp. 1--4.

\bibitem{albawi2017understanding}
S.~Albawi, T.~A. Mohammed, and S.~Al-Zawi, ``Understanding of a convolutional
  neural network,'' in \emph{2017 international conference on engineering and
  technology (ICET)}.\hskip 1em plus 0.5em minus 0.4em\relax Ieee, 2017, pp.
  1--6.

\bibitem{yu2018deep}
F.~Yu, D.~Wang, E.~Shelhamer, and T.~Darrell, ``Deep layer aggregation,'' in
  \emph{Proceedings of the IEEE conference on computer vision and pattern
  recognition}, 2018, pp. 2403--2412.

\bibitem{ma2021rw}
Y.~Ma, Z.~Ren, and S.~Xu, ``Rw-resnet: A novel speech anti-spoofing model using
  raw waveform,'' \emph{arXiv preprint arXiv:2108.05684}, 2021.

\bibitem{lee2020revisiting}
S.~Lee and C.~Lee, ``Revisiting spatial dropout for regularizing convolutional
  neural networks,'' \emph{Multimedia Tools and Applications}, vol.~79, no.~45,
  pp. 34\,195--34\,207, 2020.

\bibitem{todisco2019asvspoof}
M.~Todisco, X.~Wang, V.~Vestman, M.~Sahidullah, H.~Delgado, A.~Nautsch,
  J.~Yamagishi, N.~Evans, T.~Kinnunen, and K.~A. Lee, ``Asvspoof 2019: Future
  horizons in spoofed and fake audio detection,'' \emph{arXiv preprint
  arXiv:1904.05441}, 2019.

\bibitem{baumann2021voice}
R.~Baumann, K.~M. Malik, A.~Javed, A.~Ball, B.~Kujawa, and H.~Malik, ``Voice
  spoofing detection corpus for single and multi-order audio replays,''
  \emph{Computer Speech \& Language}, vol.~65, p. 101132, 2021.

\bibitem{loshchilov2017decoupled}
I.~Loshchilov and F.~Hutter, ``Decoupled weight decay regularization,''
  \emph{arXiv preprint arXiv:1711.05101}, 2017.

\bibitem{tak2022automatic}
H.~Tak, M.~Todisco, X.~Wang, J.-w. Jung, J.~Yamagishi, and N.~Evans,
  ``Automatic speaker verification spoofing and deepfake detection using
  wav2vec 2.0 and data augmentation,'' \emph{arXiv preprint arXiv:2202.12233},
  2022.

\bibitem{wang2021comparative}
X.~Wang and J.~Yamagishi, ``A comparative study on recent neural spoofing
  countermeasures for synthetic speech detection,'' \emph{arXiv preprint
  arXiv:2103.11326}, 2021.

\bibitem{he2015delving}
K.~He, X.~Zhang, S.~Ren, and J.~Sun, ``Delving deep into rectifiers: Surpassing
  human-level performance on imagenet classification,'' in \emph{Proceedings of
  the IEEE international conference on computer vision}, 2015, pp. 1026--1034.

\bibitem{chettri2019ensemble}
B.~Chettri, D.~Stoller, V.~Morfi, M.~A.~M. Ram{\'\i}rez, E.~Benetos, and B.~L.
  Sturm, ``Ensemble models for spoofing detection in automatic speaker
  verification,'' \emph{arXiv preprint arXiv:1904.04589}, 2019.

\bibitem{monteiro2020generalized}
J.~Monteiro, J.~Alam, and T.~H. Falk, ``Generalized end-to-end detection of
  spoofing attacks to automatic speaker recognizers,'' \emph{Computer Speech \&
  Language}, vol.~63, p. 101096, 2020.

\bibitem{gomez2019light}
A.~Gomez-Alanis, A.~M. Peinado, J.~A. Gonzalez, and A.~M. Gomez, ``A light
  convolutional gru-rnn deep feature extractor for asv spoofing detection,'' in
  \emph{Proc. Interspeech}, vol. 2019, 2019, pp. 1068--1072.

\bibitem{aravind2020audio}
P.~Aravind, U.~Nechiyil, N.~Paramparambath \emph{et~al.}, ``Audio spoofing
  verification using deep convolutional neural networks by transfer learning,''
  \emph{arXiv preprint arXiv:2008.03464}, 2020.

\bibitem{zhang2021one}
Y.~Zhang, F.~Jiang, and Z.~Duan, ``One-class learning towards synthetic voice
  spoofing detection,'' \emph{IEEE Signal Processing Letters}, vol.~28, pp.
  937--941, 2021.

\bibitem{wu2020light}
Z.~Wu, R.~K. Das, J.~Yang, and H.~Li, ``Light convolutional neural network with
  feature genuinization for detection of synthetic speech attacks,''
  \emph{arXiv preprint arXiv:2009.09637}, 2020.

\bibitem{tak2020spoofing}
H.~Tak, J.~Patino, A.~Nautsch, N.~Evans, and M.~Todisco, ``Spoofing attack
  detection using the non-linear fusion of sub-band classifiers,'' \emph{arXiv
  preprint arXiv:2005.10393}, 2020.

\bibitem{chen2020generalization}
T.~Chen, A.~Kumar, P.~Nagarsheth, G.~Sivaraman, and E.~Khoury, ``Generalization
  of audio deepfake detection.'' in \emph{Odyssey}, 2020, pp. 132--137.

\end{thebibliography}
\end{document}